\documentclass[aps,prd,showpacs,twocolumn,amsmath,10pt,superscriptaddress,floatfix,nofootinbib,a4paper]{revtex4-1}
\usepackage{epsfig,amssymb,amsfonts,bm,color}
\usepackage{tikz-feynman}
\usepackage{makecell}
\usepackage{threeparttable}
\usepackage{multirow}
\usepackage{diagbox}
\usepackage{graphicx}
\usepackage{dcolumn}
\usepackage{bm}
\usepackage{hyperref}

\usepackage{footnote}
\usepackage{tikz}
\usepackage{tikz-feynman}
\usepackage{float}
\allowdisplaybreaks[4]

\newcommand{\ket}[1]{| #1 \rangle}

\newcommand{\lra}[1]{\left(#1\right)}

\tikzfeynmanset{
every vertex/.style={red, dot},
every particle/.style={blue},
every blob/.style={draw=black!40!black, pattern color=black!40!black},
}
\usetikzlibrary{decorations.markings}
\usetikzlibrary{snakes}
\tikzset{
 photon/.style={decorate, decoration={snake}, draw=black},
    electron/.style={draw=black, postaction={decorate},
        decoration={markings,mark=at position .55 with {\arrow[draw=black]{>}}}},
    gluon/.style={decorate, draw=magenta,
        decoration={coil,amplitude=3pt, segment length=4pt}},
    scalar/.style={dashed,line width=.6pt, postaction={decorate},
        decoration={markings,mark=at position .55 with {\arrow[draw=black]{>}}}},
}

\newcommand{\itp}{\affiliation{CAS Key Laboratory of Theoretical Physics, Institute of Theoretical Physics, Chinese Academy of Sciences,\\  Zhong Guan Cun East Street 55, Beijing 100190, China}}

\newcommand{\ucas}{\affiliation{School of Physical Sciences, University of Chinese Academy of Sciences, Beijing 100049, China}}

\newcommand{\csu}{\affiliation{School of Physics, Central South University, Changsha 410083, China}}

\begin{document}

\title{Interpretation of the $\eta_1(1855)$ as a $K\bar K_1(1400)+$ c.c. molecule}

\author{Xiang-Kun Dong}
\itp\ucas

\author{Yong-Hui Lin}
\affiliation{Helmholtz-Institut~f\"{u}r~Strahlen-~und~Kernphysik~and~Bethe~Center~for
Theoretical~Physics, \\Universit\"{a}t~Bonn,  D-53115~Bonn,~Germany}

\author{Bing-Song Zou}
\email{zoubs@itp.ac.cn}
\itp\ucas\csu

\begin{abstract}
An exotic state with $J^{PC}=1^{-+}$,
denoted by $\eta_1(1855)$, was observed by BESIII collaboration recently in $J/\psi \to \gamma \eta\eta'$. 
The fact that its mass is just below the threshold of $K\bar K_1(1400)$ stimulates us to investigate whether this exotic state can be interpreted as a $K\bar K_1(1400)+$ c.c { molecule or not}.
Using the one boson exchange model, we show that it is possible for $K\bar K_1(1400)$ with $J^{PC}=1^{-+}$ to bind together by taking the momentum cutoff $\Lambda\gtrsim 2$ GeV and yield the same binding energy as the experimental value when $\Lambda\approx 2.5$ GeV. In this molecular picture, the predicted branch ratio $\mathrm{Br}(\eta_1(1855)\to\eta\eta') \approx 15\%$ is consistent with the experimental results, which again supports the molecular explanation of $\eta_1(1855)$. Relevant systems, namely $K\bar K_1(1400)$ with $J^{PC}=1^{--}$ and $K\bar K_1(1270)$ with $J^{PC}=1^{-\pm}$, are also investigated, some of which can be searched for in the future experiments.
\end{abstract}


\maketitle

\section{Introduction}
Many efforts have been put into searching for non-conventional hadrons other than quark-antiquark mesons or 3-quark baryons since the quark model was proposed by Gell-Mann~\cite{GellMann:1964nj} and Zweig~\cite{Zweig:1964jf}.  Quantum Chromodynamics (QCD), the most fundamental theory describing the strong interaction, does not forbid the existence of multi-quark states, hybrid states or glueballs, collectively called exotic hadrons. Actually, there were several candidates for exotic hadrons before the beginning of this century but none of them are identified unambiguously~\cite{Klempt:2007cp}. For example, even now there is not a widely accepted explanation for the nature of $\Lambda(1405)$, which was ever discovered almost 60 years ago~\cite{Dalitz:1959dn,Dalitz:1960du,Alston:1961zzd}, see the discussions in the most recent reviews~\cite{Hyodo:2020czb,Mai:2020ltx}.

Since the discovery of the charmonium-like state X(3872)~\cite{Choi:2003ue}, the last twenty years has witnessed the booming experimental evidence of exotic states, see Refs.~\cite{Chen:2016qju,Hosaka:2016pey,Richard:2016eis,Lebed:2016hpi,Esposito:2016noz,Guo:2017jvc,Ali:2017jda,Olsen:2017bmm,Kou:2018nap,Kalashnikova:2018vkv,Cerri:2018ypt,Liu:2019zoy,Brambilla:2019esw,Guo:2019twa,Yang:2020atz,Ortega:2020tng,Dong:2021juy,Dong:2021bvy} for recent reviews. To identify exotic states or their candidates, the following guidelines may be helpful: a) In the charmonium and bottomonium sectors, quark models give quite precise descriptions of the conventional mesons (see, e.g., Refs.~\cite{Godfrey:1985xj,Capstick:1985xss}). If the observed resonances do not fit such properties, they can be considered as candidates of exotic states, such as X(3872) and Y(4260)~\cite{BaBar:2005hhc}; b) In some cases, the decay channels imply the resonances consist of multiquarks and therefore, these states are exotic definitely, such as $Z_{c(s)}$ states~\cite{Ablikim:2013mio,Liu:2013dau,BESIII:2020qkh}, $P_{c(s)}$ states~\cite{Aaij:2015tga,Aaij:2019vzc,LHCb:2020jpq} and most recently observed doubly charmed $T_{cc}$ state~\cite{LHCb:2021vvq,LHCb:2021auc}; c) Some combinations of $J^{PC}$ for conventional mesons are forbidden, such as $0^{--},(\rm{even})^{+-}$ and $(\rm{odd})^{-+}$. Therefore, a state with such forbidden quantum numbers must be exotic. Up to now there are two well-established states with exotic $J^{PC}$ in light quark sector, $\pi_1(1400),\pi_1(1600)$~\cite{ParticleDataGroup:2020ssz} and one possible state $\pi_1(2015)$~\cite{Kuhn:2004en,Lu:2004yn}\footnote{See Ref.~\cite{Meyer:2015eta} for a recent review on the experimental status of these $\pi_1$'s.}, all of which have $J^{PC}=1^{-+}$, while no experimental signals appear in the heavy quark sector.

Recently, BESIII Collaboration~\cite{BESIII:2022PRL,BESIII:2022PRD} reported a resonance, denoted as $\eta_1(1855)$, with $m=1855\pm 9^{+6}_{-1}$ MeV and $\Gamma=188\pm 18^{+3}_{-8}$ MeV, in the invariant mass distribution of $\eta\eta'$ in $J/\psi \to \gamma \eta\eta'$ with a significance of 19 $\sigma$. Its quantum numbers $J^{PC}=1^{-+}$ ensure it to be an exotic state, {which immediately stimulates the explanation of $\eta_1(1875)$ being an isoscalar hybrid candidate~\cite{Chen:2022qpd,Qiu:2022ktc}}. Notice that this state locates at just around 40 MeV below the $K\bar K_1(1400)$ threshold and therefore, it is natural to interpret it as a molecule of $K\bar K_1(1400)+$ c.c.($K\bar K_1$ for short in the following), which can decay into $\eta\eta'$ via $K^*$ exchange on the hadronic level. 

In this paper, we investigate whether $K\bar K_1$ can form a bound state via the interaction driven by one boson exchange as shown in Fig.~\ref{fig:Feyn_KK1}. Pseudoscalar exchange between $K\bar K_1$ is forbidden by parity conservation. Hence it is expected that the $\sigma$ and vector meson ($\rho,\omega,\phi$) exchanges dominate the interaction. In the Review of Particle Physics (RPP)~\cite{ParticleDataGroup:2020ssz} there are two $K_1$'s, $K_1(1270)$ with mass 1270 MeV and $K_1(1400)$ with mass 1403 MeV. These two mass eigenstates have quite different decay behaviors and are considered as the mixing of two flavor eigenstates from the $^3P_1$ and $^1P_1$ octets~\cite{Burakovsky:1997dd, Suzuki:1993yc, Cheng:2003bn, Yang:2010ah, Hatanaka:2008xj, Tayduganov:2011ui, Divotgey:2013jba, Zhang:2017cbi}. It was also explored in Refs.~\cite{Roca:2005nm, Geng:2006yb, Wang:2019mph} that the $K_1(1270)$ may have a two-pole structure in vector-pseudoscalar scattering. In this paper we adopt the former treatment to investigate their interactions. 

With the above considerations we find that a) the meson exchange potential with reasonable cutoff is strong enough for $K\bar K_1(1400)$ to form a bound state; b) The branch ratio of the predicted molecule decaying into $\eta\eta'$ is consistent with the experimental data of BESIII~\cite{BESIII:2022PRL,BESIII:2022PRD}, Br$(J / \psi \rightarrow \gamma \eta_{1}(1855) \rightarrow \gamma \eta \eta^{\prime})=\left(2.70 \pm 0.41_{-0.35}^{+0.16}\right) \times 10^{-6}$; c) Its $C-$parity partner are predicted to have a similar binding energy and the $\pi\rho$, $\eta\omega$ and $K\bar K$ channels are good places to search for it. The properties predicted in this paper serve as guidance for further experimental explorations of these states and future experiments can in turn test our molecular state assignment for the observed $\eta_1(1855)$.

\section{Possible bound states of $K\bar K_1(1400)$}

Due to the $U(3)_V$ flavor symmetry breaking effect derived from the mass difference between $u/d$ and $s$ quarks, the axialvector $K_{1A}$ ($^3P_1$ state) and pseudovector $K_{1B}$ ($^1P_1$ state) can mix with each other and generate the two physical resonances $K_1(1270)$ and $K_1(1400)$. Following Ref.~\cite{Divotgey:2013jba} the mixing is parameterized as

\begin{align}
    \left(\begin{array}{cc}{\left|K_{1}(1270)\right\rangle}  \\ {\left|K_{1}(1400)\right\rangle}\end{array}\right)=\left(\begin{array}{cc}{\cos \theta} & {-i\sin \theta}\\ {-i\sin \theta} & {\cos \theta} \end{array}\right)\left(\begin{array}{c}{\left|K_{1A}\right\rangle} \\ {\left|K_{1B}\right\rangle}\end{array}\right)
\end{align}
where the mixing angle $\theta$ is determined to be $(56.4\pm4.3)^{\circ}$. The flavor wave functions of $\ket{K\bar K_1}$ with positive and negative C-parity read
\begin{align}
C=\pm:\frac{1}{\sqrt 2}\lra{\ket{K\bar K_1}\pm\mathcal C\ket{ KK_1}}
\end{align}
with the following conventions, $\mathcal C\ket{K}=\ket{\bar K}$, $\mathcal C\ket{K_{1A}}=\ket{\bar K_{1A}}$ and $\mathcal C\ket{K_{1B}}=-\ket{\bar K_{1B}}$. Here $\mathcal C$ is the charge conjugation operator.
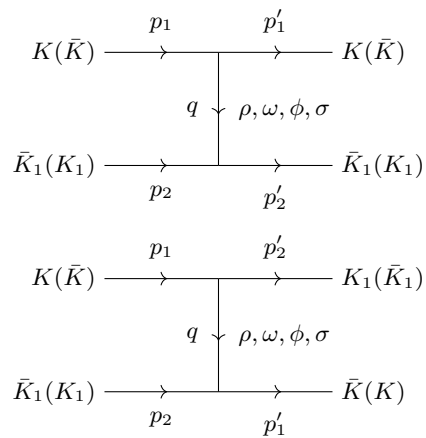
\begin{figure}
\centering
\begin{tikzpicture}[scale=1.5]
\draw[electron] (-1,0.5)--(0,0.5);
\draw[electron] (0,0.5)--(1,0.5);
\draw[electron] (-1,-0.5)--(0,-0.5);
\draw[electron] (0,-0.5)--(1,-0.5);
\draw[electron] (0,0.5)--(0,-0.5);
\node[right] at (0.1,0){$\rho,\omega,\phi,\sigma$};
\node[left] at (-1,0.5){$K(\bar K)$};
\node[left] at (-1,-0.5){$\bar K_1(K_1)$};
\node[right] at (1,0.5){$K_1(\bar K_1)$};
\node[right] at (1,-0.5){$\bar K(K)$};
\node[left] at (-0.1,0){$q$};
\node [above] at (-0.5,0.6){$p_1$};
\node [above] at (0.5,0.6){$p'_2$};
\node [below] at (-0.5,-0.6){$p_2$};
\node [below] at (0.5,-0.6){$p'_1$};

\begin{scope}[yshift=2cm]

\draw[electron] (-1,0.5)--(0,0.5);
\draw[electron] (0,0.5)--(1,0.5);
\draw[electron] (-1,-0.5)--(0,-0.5);
\draw[electron] (0,-0.5)--(1,-0.5);
\draw[electron] (0,0.5)--(0,-0.5);
\node[right] at (0.1,0){$\rho,\omega,\phi,\sigma$};
\node[left] at (-1,0.5){$K(\bar K)$};
\node[left] at (-1,-0.5){$\bar K_1(K_1)$};
\node[right] at (1,0.5){$K(\bar K)$};
\node[right] at (1,-0.5){$\bar K_1(K_1)$};
\node[left] at (-0.1,0){$q$};
\node [above] at (-0.5,0.6){$p_1$};
\node [above] at (0.5,0.6){$p'_1$};
\node [below] at (-0.5,-0.6){$p_2$};
\node [below] at (0.5,-0.6){$p'_2$};
\end{scope}

\end{tikzpicture}

\caption{Feynman diagrams for direct (top) and cross (bottom) meson exchange between $K\bar K_1$. }\label{fig:Feyn_KK1}
\end{figure}

Working with the chiral perturbation theory, the pseudoscalar-pseudoscalar-vector coupling is described by~\cite{Meissner:1987ge}
\begin{align}
\mathcal{L}_{\rm{PPV}}&=i\sqrt{2}\,G_{\rm V}\,{\rm{Tr}}\lra{[\partial_{\mu}P,P]V^{\mu}}\label{eq:Lagran_PPV}
\end{align}
and analogously, we assume the following couplings
\begin{align}
\mathcal{L}_{{AAV}}&=i\sqrt{2}\,G_{\rm V}'\,{\rm{Tr}}\lra{[\partial_{\mu}A^\nu,A_\nu]V^{\mu}}\label{eq:Lagran_AAV}\\
\mathcal{L}_{{BBV}}&=i\sqrt{2}\,G_{\rm V}'\,{\rm{Tr}}\lra{[\partial_{\mu}B^\nu,B_\nu]V^{\mu}}\label{eq:Lagran_BBV}\\
\mathcal{L}_{APV}&= i a\ {\rm Tr}(A_\mu[V^\mu,P]),\\
\mathcal{L}_{BPV}&= b\ {\rm Tr}(B_\mu\{V^\mu,P\})
\end{align}
with 
\begin{align}
P=\left(\begin{array}{ccc}
\frac{\pi^{0}}{\sqrt{2}}+\frac{\eta}{\sqrt{6}} & \pi^{+} & K^{+} \\
\pi^{-} & -\frac{\pi^{0}}{\sqrt{2}}+\frac{\eta}{\sqrt{6}} & K^{0} \\
K^{-} & \bar{K}^{0} & -\sqrt{\frac{2}{3}} \eta
\end{array}\right),
\end{align}
\begin{align}
V=\left(\begin{array}{ccc}
\frac{\omega}{\sqrt{2}}+\frac{\rho^{0}}{\sqrt{2}} & \rho^{+} & K^{*+} \\
\rho^{-} & \frac{\omega}{\sqrt{2}}-\frac{\rho^{0}}{\sqrt{2}} & K^{* 0} \\
K^{*-} & \bar{K}^{* 0} & \phi
\end{array}\right),
\end{align}
and 
\begin{align}
A(B)=\left(\begin{array}{ccc}{*} & {*} & {K_{1A(B)}^{+}} \\ {*} & {*} & {K_{1A(B)}^{0}} \\ {K_{1A(B)}^{-}} & {\bar{K}_{1A(B)}^{0}} & *\end{array}\right)\label{eq:axialnonet}.
\end{align}
Here all irrelevant axialvector and pseudovector states are labeled as ``*" in the multiplet matrix Eq.~\eqref{eq:axialnonet}. The coupling constant $G_{\rm V}\approx3.0$ was estimated from the decay width of $\rho\to\pi\pi$~\cite{Zhang:2006ix} and we assumed the magnitude of $G^\prime_V$ is comparable with $G_{\rm V}$, i.e., $|G'_{\rm V}|\approx G_{\rm V}$. Moreover, taking the mixing angle obtained in Ref.~\cite{Divotgey:2013jba} and fitting to the experimental widths of $K_1(1270)$/$K_1(1400)$ decaying to $K\rho$, one can determine the coupling constants as $a\approx1.92\pm0.09$ GeV and $b\approx-2.47\pm0.08$ GeV~\cite{Dong:2020rgs}.

In our previous work~\cite{Dong:2019ofp}, we have discussed the $D\bar D_1+$ c.c. bound state via one boson exchange to understand the nature of Y(4260). $KK_1$ system behaves similarly with $DD_1$ system in the one boson exchange picture, just replacing the $c$ quarks in $DD_1$ with $s$ quarks, except that $KK_1$ can couple to $\phi$ while $DD_1$ can not. Therefore one can expect that they have the same Lagrangian except the slightly different values of coupling constants. Note that the $DDV$ coupling constant is 0.9 of the $KKV$ coupling~\cite{Casalbuoni:1996pg,Isola:2003fh}. These observations make us confident that the Language in Eqs.~(\ref{eq:Lagran_AAV}, \ref{eq:Lagran_BBV}) can give a good description of the $K_1K_1V$ vertices and also indicate the following couplings,
\begin{align}
    \mathcal{L}_{KK \sigma}&=-2g_{\sigma}m_K\bar{K}K \sigma,\\
    \mathcal{L}_{K_1K_1 \sigma}&=-2g''_{\sigma}m_{K_1}\bar{K^\mu_1}K_{1\mu} \sigma,\label{eq:K1K1sigma}\\
    \mathcal{L}_{K_1K \sigma}&=-\frac{2 \sqrt{6}}{3} \frac{h_{\sigma}^{\prime}}{f_{\pi}} \sqrt{m_K m_{K_1}}\left(\bar{K}_{1}^{\mu} K+\bar{K} K_{1}^{\mu}\right)\partial_{\mu} \sigma\label{eq:K1Ksigma}
\end{align}
with $f_\pi= 132\ \rm{MeV}$ the pion decay constant. As a rough approximation, we take $g_\sigma=g''_\sigma=-0.76$, $h'_\sigma=0.35$~\cite{Bardeen:2003kt}. We have ignored the mixing nature of $K_1(1270)$ and $K_1(1400)$ in Eqs.(\ref{eq:K1K1sigma}, \ref{eq:K1Ksigma}), which does not matter here.

The potential of $K\bar K_1$ in the non-relativistic limit reads
\begin{equation}
    \tilde V(\bm q)=-\frac{\mathcal M}{4m_K m_{K_1}}
\end{equation}
with $\mathcal M$ the invariant scattering amplitude of $K\bar K_1\to K\bar K_1$. The potentials in momentum space for the direct and cross scatterings, see the top and bottom diagram in Fig.~\ref{fig:Feyn_KK1}, are written as
\begin{align}
\tilde V_{\rm V,d}({\bf q})&=f_{\mathrm I}g_{\rm {V,d}}g'_{\rm {V,d}}\frac{1}{|{\bf q}|^2+m_{\rm V}^2}\label{eq:Poten_Momentum_t},\\
\tilde V_{\rm {V,c}}({\bf q})&=\frac{f_{\mathrm I}g^2_{\rm {V,c}}}{4m_{\rm{K}}m_{{\rm{K}_1}}}\frac{\lra{1+{\bm\epsilon}_1\cdot \bm{q\epsilon}_2^*\cdot \bm{q}/m_{\rm{V}}^2}}{|{\bm q}|^2+\tilde{m}_{\rm V}^2}\label{eq:Poten_Momentum_u},\\
\tilde V_{\sigma, \rm{d}}({\bf q})&=-g_\sigma g''_{\sigma}\frac{1}{|{\bf q}|^2+m_{\sigma}^2},\\
\tilde V_{\sigma, \rm{c}}({\bf q})&=\frac{-2h^{'2}_\sigma}{3f_{\pi}^2}\frac{{\bm\epsilon}_1\cdot \bm{q\epsilon}_2^*\cdot \bm{q}}{|{\bm q}|^2+\tilde{m}_{\sigma}^2}\label{eq:Poten_Momentum_u_sigma}
\end{align}
with $\tilde m_{\rm{V}/\sigma}^2=m_{\rm V/\sigma}^2-(m_{\rm{K}_1}-m_{\rm{K}})^2$. 
The coupling constants are
\begin{align}
g_{\rho,\rm d}&=g_{\omega,\rm d}=-\frac1{\sqrt{2}}g_{\phi,\rm d}=G_{\rm V}\\
g'_{\rho,\rm d}&=g'_{\omega,\rm d}=-\frac1{\sqrt{2}}g'_{\phi,\rm d}=G'_{\rm V}\\
g^2_{\rho,\rm c}&=g^2_{\omega,\rm c}=\frac1{2}g^2_{\phi,\rm c}=\frac12(a\sin\theta\pm b\cos\theta)^2
\end{align}
with ``$+$" for $K_1(1400)$ and ``$-$" for $K_1(1270)$ and the isospin factors, $f_{\mathrm I}$, are 
\begin{align}
f_{0}=\left\{\begin{array}{cc}{3} & {\rho} \\ {1} & {\omega} \\ {1} & {\phi}\end{array}\right.\ \ \ \ f_{1}=\left\{\begin{array}{cc}{-1} & {\rho} \\ {1} & {\omega} \\ {1} & {\phi}\end{array}\right..
\end{align}

A form factor should be introduced at each vertex to account for the finite size of the involved mesons. Here we take the commonly used monopole form factor,
\begin{equation}
F(q,m,\Lambda)=\frac{\Lambda^2-m^2}{\Lambda^2-q^2},\label{eq:form_factor}
\end{equation}
which in position space can be looked upon as a spherical source of the exchanged mesons~\cite{Tornqvist:1993ng}. The potentials in position space can be obtained by Fourier transformation of Eqs.(\ref{eq:Poten_Momentum_t}, \ref{eq:Poten_Momentum_u}), together with the form factor Eq.~(\ref{eq:form_factor}),

\begin{align}
V_{\rm {V,d}}\left(\mathbf{r}\right)&=f_{\mathrm I}g_{\rm{V,d}}g'_{\rm{V,d}}{\Big(}m_{\rm{V}}Y(m_{\rm{V}} r) -\Lambda Y(\Lambda r)\notag\\
&\ \ \ \ -\frac{1}{2}(\Lambda^{2}-m_{\rm{V}}^{2})r Y(\Lambda r)\Big),\\
V_{\rm {V,c}}\left(\mathbf{r}\right)&=\frac{f_{\mathrm I}g^2_{\rm{V,c}}}{4m_{\rm{K}}m_{{\rm{K}_1}}}\Big\{\Big[\tilde{m}_{\rm{V}}Y(\tilde{m}_{\rm{V}} r)-\tilde\Lambda Y(\tilde\Lambda r)\notag\\
&\ \ \ \ -\frac{1}{2}(\tilde\Lambda^{2}-\tilde m^{2}_{\rm{V}})r Y(\tilde\Lambda r)\Big]\notag\\
&\ \ \ \ -\frac{1}{3m_{\rm{V}}^2}\Big[\tilde m^2_{\rm{V}}\lra{\tilde{m}_{\rm{V}}Y(\tilde{m}_{\rm{V}} r)-\tilde\Lambda Y(\tilde\Lambda r)}\notag\\
&\ \ \ \ -\frac{1}{2}\lambda^2(\tilde\Lambda^{2}-\tilde m^{2}_{\rm{V}})r Y(\tilde\Lambda r)\Big]
\Big\}\label{eq:VVc}\\
V_{\rm {\sigma,d}}\left(\mathbf{r}\right)&=-g_\sigma g''_{\sigma}\Big(m_{\sigma}Y(m_{\sigma} r) -\Lambda Y(\Lambda r)\notag\\
&\ \ \ \ -\frac{1}{2}(\Lambda^{2}-m_{\sigma}^{2})r Y(\Lambda r)\Big),\\
V_{\sigma,\rm {c}}\left(\mathbf{r}\right)&=\frac{2h^{'2}_\sigma}{9f_{\pi}^2}\Big\{\Big[\tilde m^2_{\sigma}\lra{\tilde{m}_{\sigma}Y(\tilde{m}_{\sigma} r)-\tilde\Lambda Y(\tilde\Lambda r)}\notag\\
&\ \ \ \ -\frac{1}{2}\lambda^2(\tilde\Lambda^{2}-\tilde m^{2}_{\sigma})r Y(\tilde\Lambda r)\Big]
\Big\}\label{eq:Vsigmac}
\end{align}
where $Y(x)=e^{-x}/4\pi x$ and $\tilde \Lambda^2=\Lambda^2-(m_{\rm K_1}-m_{\rm K})^2.$ Note that the Fourier transformation of Eqs.(\ref{eq:Poten_Momentum_u},\ref{eq:Poten_Momentum_u_sigma}) contains a $\delta$ function, which is omitted in some works, see the discussion in Ref.~\cite{Burns:2019iih}. The $\delta$ potential is from short distance physics, which we are blind to, and may contain contributions from the exchange of heavier mesons not considered here. In Ref.~\cite{Yalikun:2021bfm}, the $\delta$ contribution is varied in the range of $0\sim 1$ to take into account the uncertainty of the unknown short distance interaction in the formation of $P_c$ states. Here we consider the marginal cases where the $\delta$ function is kept and omitted, corresponding to $\lambda=\tilde\Lambda$ and $\lambda=\tilde m_{\rm V}$ in Eqs.~(\ref{eq:VVc}, \ref{eq:Vsigmac}), respectively. We take the real part of potentials when they are complex. The total potential for positive or negative C-parity $K\bar K_1$ system reads
\begin{align}
    V^{{I},C=\pm}=\left(V_{\rm V,d}^{I}+V_{\sigma,\rm d}\right)\pm \left(V_{\rm V,c}^{I}+ V_{\sigma, \rm c}\right)\label{eq:potentialCpm}
\end{align}
with $I=0,1$. We take the following masses from RPP~\cite{ParticleDataGroup:2020ssz} to perform numerical calculations, $m_{{K}}=0.495\ \rm{GeV}$, $m_{{K_1}(1400)}=1.403\ \rm{GeV}$, $m_{{K_1}(1270)}=1.272\ \rm{GeV}$, $m_\rho=0.775\ \rm{GeV}$, $m_{\omega}=0.783\ \rm{GeV}$, $m_\phi=1.019\ \rm{GeV}$ and $m_\sigma=0.600\ \rm{GeV}$.

\begin{figure}[h]
    \centering
    \includegraphics[width=\linewidth]{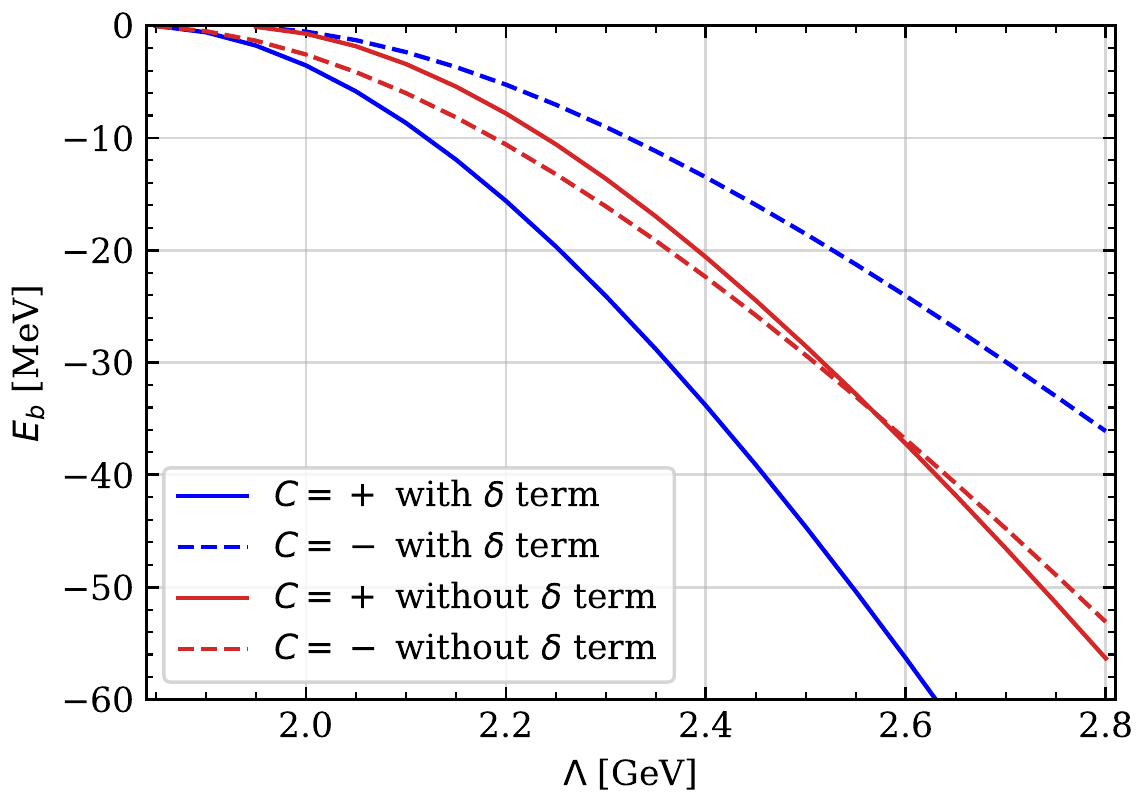}
    \caption{The binding energies of the isoscalar $K\bar K_1(1400)$ bound states.}\label{fig:Eb1400}
\end{figure}

The Schr\"odinger equations for both $K\bar K_1(1400)$ and $K\bar K_1(1270)$ systems are solved numerically. For the isovector cases, no bound states are found since contributions from $\rho$ and $\omega$ exchanges cancel with each other. Now we focus on the isoscalar cases. The binding energies for $K\bar K_1(1400)$ with different cutoffs are shown in Fig.~\ref{fig:Eb1400}. We can see that the $1^{-+}$ system can form a bound state when $\Lambda\gtrsim 1.9$ GeV. We need $\Lambda\approx 2.5$ GeV to produce a binding energy of 40 MeV, the experimental value of the newly observed exotic state by BESIII. With the same cutoff, the binding energy of the $1^{--}$ $K\bar K_1(1400)$ state is predicted to be $10\sim30$ MeV where the uncertainty comes from whether the $\delta$ potential is included or not. Due to the large coupling of $K_1(1270)KV$, the potential between $K\bar K_1(1270)$ is repulsive for $C=+$ and therefore, no bound states are expected. On the other hand, the potential is too attractive for $C=-$ to form a molecule with a reasonable binding energy if we use a cutoff about $2.5$ GeV. Even when $\Lambda\sim1.5$ GeV the binding energy is larger than 100 MeV, which is not acceptable for a shallow bound state, for which our previous treatments hold valid.

\section{Decay properties of $K\bar K_1$ molecules}

\begin{figure}[h]
\centering
\includegraphics[width=\linewidth]{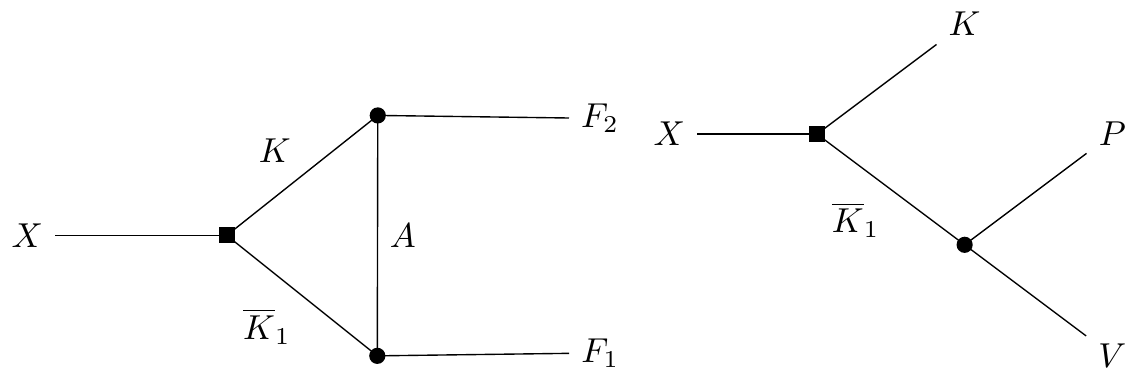}
\caption{Two-body and three-body decays of molecules composed of $K\bar K_1$.}\label{fig:decayfeyn}
\end{figure}

With reasonable cutoff we have obtained the possible bound states of $K\bar K_1(1400)$ with $J^{PC}=1^{-\pm}$, denoted as $X$ temporarily. It is now desirable to estimate the decay patterns of the predicted molecular states, especially the $\eta\eta'$ channel, to compare with BESIII observation and provide guidance to further experimental investigations. It is natural that such molecules decay through their components, as illustrated in Fig.~\ref{fig:decayfeyn}. Since $K_1$'s have a quite large decay width, we assume that the three-body decays of the molecules are dominated by the process shown in the right diagram in Fig.~\ref{fig:decayfeyn}, where $P,V=\pi, K^*$ or $K,\rho/\omega$. Two-body strong decay channels considered here are listed in Table~\ref{tab:widths}.

The coupling between a hadronic molecule and its components can be constructed as
\begin{align}
\mathcal L_{XKK_1}=y X^{\mu}K^\dagger K^\dagger_{1\mu}+ h.c.
\end{align}
where $y$ is the coupling constant and $X$ denotes the field of the molecule. For a loosely bound state, one can estimate the coupling constant in a model-independent way by means of Weinberg compositeness criterion~\cite{Weinberg:1965zz,Baru:2003qq,Guo:2017jvc}, which is developed to estimate the amount of compact and molecular components in a given state, where the molecular components consist of two stable (or narrow~\footnote{``narrow" is relative to the binding energy of that molecular state, that is, the widths of the molecular components should be much smaller than its binding energy.}) compact states. However, $K_1(1400)$ has a width of 174 MeV, much larger than the obtained binding energy in our case. Strictly speaking, one should consider the dynamics between the sub-ingredients of $K_1(1400)$ and kaon, such as $K \bar{K}^* \pi$ three-body system, like what were done for $X(3872)$~\cite{Baru:2011rs} and $T_{cc}^+$~\cite{Du:2021zzh}. {On the other hand, the decay width of $K_1(1400)$ is much smaller than the masses of exchanged $\rho$ and $\omega$, and hence it should not jeopardize the molecular interpretation of $\eta_1(1855)$ as 
argued in Ref.~\cite{Guo:2011dd}.}

To obtain a better estimate of the coupling $y$, we consider in the following the pole position of the $X$ on the complex $\sqrt{s}$ plane by assuming that the $X$ is a purely molecular state of $K\bar K_1(1400)$. The dominant decay channel of $K_1(1400)$ is $K^*\pi$ with a branching ratio of 94\% and for simplicity, we take it to be 100\% by ignoring all other channels. The pole position of $K\bar K(1400)$ system with coupled channel effects ignored is determined by
\begin{equation}
    1-VG=0
\end{equation}
where $V$, assumed to be constant, is the interaction strength of $K\bar K(1400)$ and $G$ is the scalar two particle loop propagator,
\begin{align}
    G(s)=\int\frac{{\rm d}^4l}{(2\pi)^4} \frac{1}{l^2-m_K^2+i\epsilon}\frac{1}{(P-l)^2-m_{K_1}^2+im_{K_1}\Gamma_{K_1}}.
\end{align}
with $P$ the total momentum. {Note that we have ignored the spin structures of the intermediate $K_1$, the corrections from which, as argued in Refs~\cite{Roca:2005nm,Molina:2008jw}, are at the order of $l_{\rm on}^2/m_{K_1}^2$ with $l_{\rm on}$ the on-shell 3-momentum of $K\bar K_1$.} A hard cutoff, $\Lambda$, of the 3-momentum $\bm l$ will be introduced to regularize the UV divergence. The running width of $K_1$ reads
\begin{align}
    \Gamma_{K_1}(s)=g_{K_1}^2\frac{|{\bm q}|}{16\pi m_{K_1}^2}\left(3+\frac{{\bm q}^2}{m_{K^*}^2}\right)
\end{align}
with $g_{K_1}=3.65$ GeV the coupling constant of $K_1K^*\pi$ and $\bm q$, depending on $s$, the 3-momentum of $K^*$ in the center of mass frame of $K_1$. 

\begin{figure}[htbp]
\centering
\includegraphics[width=\linewidth]{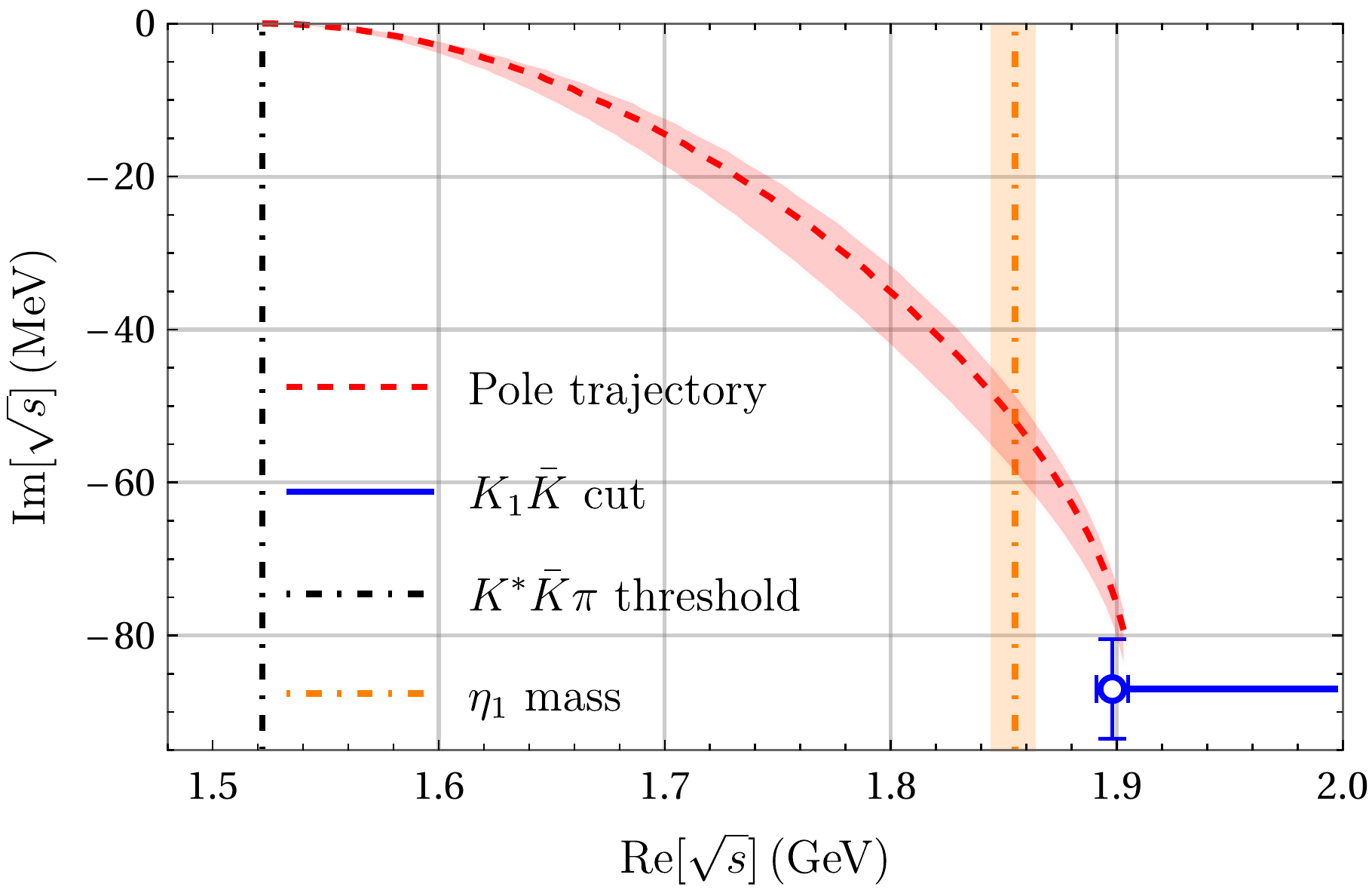}
    \caption{Pole trajectory of the $K\bar{K}_1$ molecule in the $S$-wave $K\bar{K}_1$ single channel model. Red line shows the pole trajectory { with $\Lambda=0.8$ GeV} and red shadow is the uncertainties caused by varying the hard momentum cut from 0.6~GeV to 1.0~GeV. Open blue circle denotes the branch point corresponding to the $K\bar{K}_1$ threshold and the blue line indicates the corresponding branch cut. The vertical black dot-dashed line is the $K \bar{K}^* \pi$ three-body threshold. The orange band shows the observed $\eta_1$ mass.}\label{fig:pole}
\end{figure}

The pole trajectory when varying $V$ is shown in Fig.~\ref{fig:pole}. As expected, the partial width of $\eta_1(1855)\to K(\bar K_1(1400)\to \bar K^*\pi)$ gets smaller when the binding energy becomes larger and the pole will touch the $K\bar K_1(1400)$ cut, around $1900-i87$ MeV, if the binding energy goes to zero. For the measured $\eta_1$ mass, the three body decay width of $X$ is around $105$ MeV, which weakly depends on the hard cutoff $\Lambda$ in the phenomenological reasonable range of $0.6\sim 1$ GeV. By matching the three body decay width of $X$ to the obtained pole position, we have an estimate of the coupling constant,
\begin{equation}
    y=13.6\ \rm GeV.
\end{equation}

We refer to our previous work for other relevant couplings~\cite{Dong:2020rgs}. In particular, a non-zero $D$-wave component is introduced in the vertex $K_1K^*\pi$ and $K_1K^*\eta$ as similar with $D_1D^*\pi$ interaction used in Ref.~\cite{Dong:2020rgs}. The ratio of $D$-wave component over the $S$-wave from  RPP~\cite{ParticleDataGroup:2020ssz} is $0.04\pm0.01$. When performing the loop integral in the triangle diagram of the two-body decays, a Gaussian regulator $f_1$ (at the vertex of $XK\bar K_1$) and a multipole form factor $f_2$ (at one of the vertices where the exchanged particle $A$ is involved), formulated as follows, are included to regularize the ultraviolet divergence.
\begin{align}
&f_1(\bm{p}^2 /\Lambda_0^2) = {\rm{exp}}(-\bm{p}^2 /\Lambda_0^2),\label{eq:gaussian}\\
&f_2(q^2) = \frac{\Lambda_1^4}{(m^2 - q^2)^2 + \Lambda_1^4}, \label{eq:multipolar}
\end{align}
where $\bm{p}$ are the three dimensional space part of momentum of $K$ or $\bar K_1$, $m$ and $q$ is the mass and momentum of the exchanged particle.

\begin{table}[htbp]
	\centering
	\caption{\label{tab:widths}Partial widths of the isoscalar $K\bar K_1(1400)$ molecular states with quantum numbers $1^{--}$ and $1^{-+}$ with $\Lambda_0=1.3$~GeV and $\Lambda_1=2.5$~GeV. All the decay widths are in unit of $\mathrm{MeV}$. Here, we assume $5\%$ $D$-wave contribution in the $K_1K^*\pi$ and $K_1K^*\eta$ vertex. }
	\scalebox{0.9}{
		\begin{tabular}{c|*{6}{c}}
			\Xhline{1pt}
			\multirow{2}*{Mode} & \multicolumn{2}{c}{Widths ($\mathrm{MeV}$)} \\
			\Xcline{2-3}{0.4pt}
			& {$1^{--}~E_B=20~\mathrm{MeV}$} & {$1^{-+}~E_B=40~\mathrm{MeV}$} \\
			\Xhline{0.8pt}
			$K^*\bar{K}^*$   & 38.1 &26.3\\
			$K \bar{K}$  	 & 0.5  & 0\\
			$K \bar{K}^*$ 	 & 1.0 & 0.9\\
			$a_1 \pi$ 	 & 0 & 9.2\\
			$f_1 \eta$ 	 & 0 & 0.2\\
			$\eta \eta^{\prime}$ 	 & 0 &26.9\\
			$\sigma \omega$ & 0.2 &0\\
			$\rho \rho$ 		 & 0 &0.04\\
			$\pi \rho$ 		 & 6.4 &0\\
			$\eta \omega$ 		 & 0.4 &0\\
			$\omega \omega$ 		 & 0 &0.01\\
			$\omega\phi$& 0 &0.4\\
			$K \bar{K}^* \pi$    & 130.0 & 105.0 \\
			\Xhline{0.8pt}
			2-body & 46.5  & 64.0\\
			Total& 176.5  & 169.0\\
			\Xhline{1pt}
	\end{tabular}}
\end{table}

\begin{figure}[htbp]
\centering
\includegraphics[width=\linewidth]{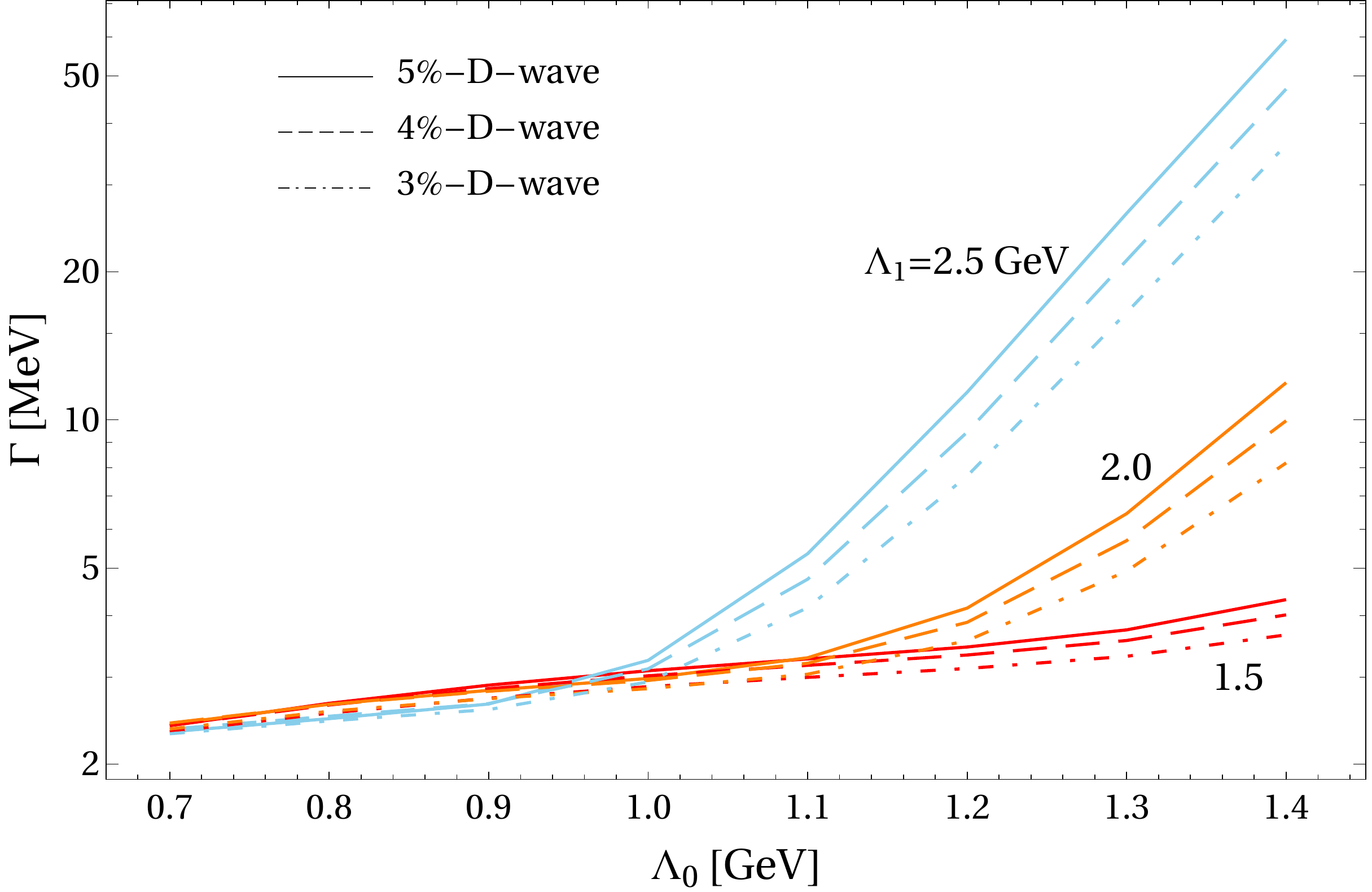}
\includegraphics[width=\linewidth]{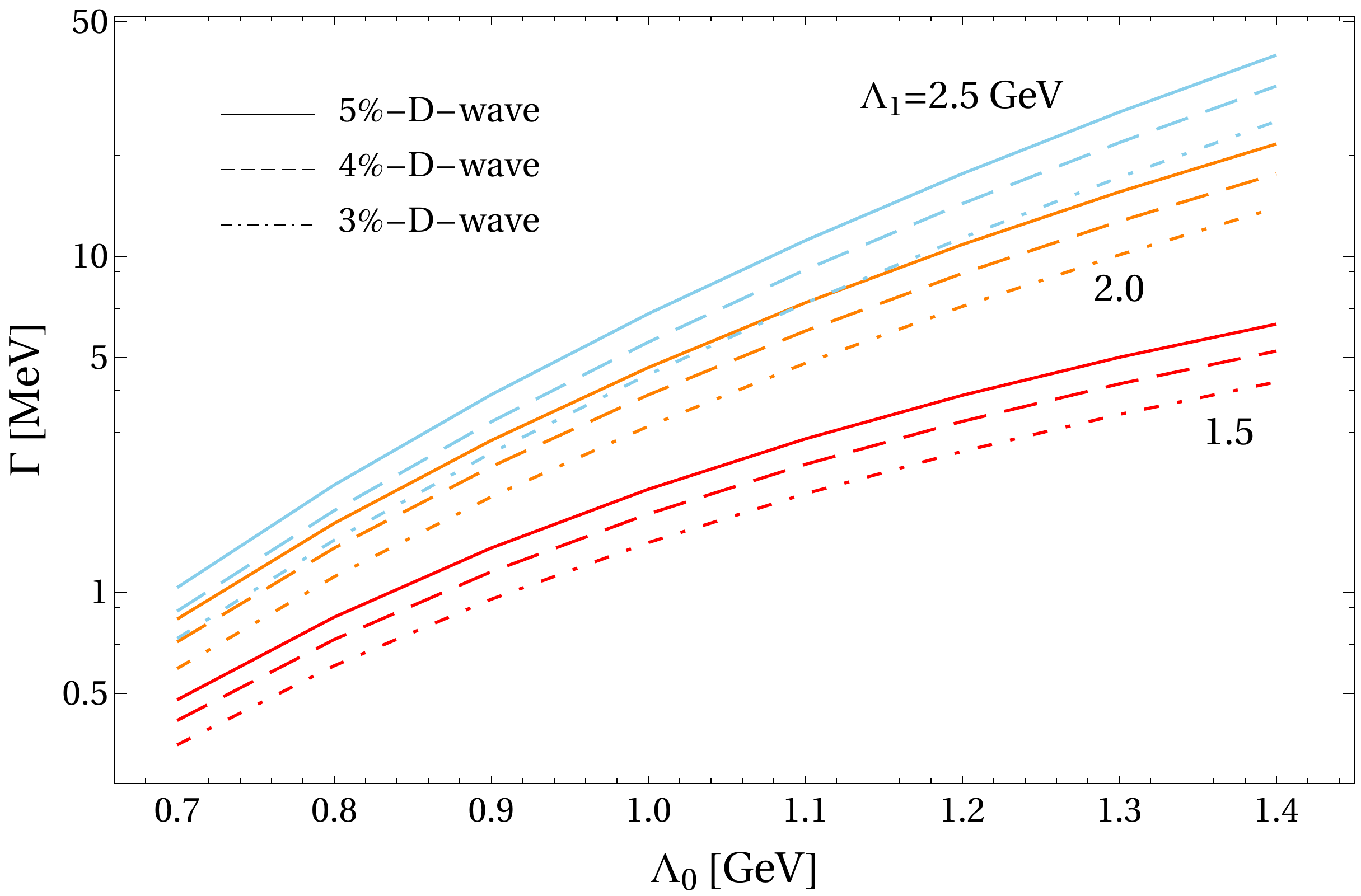}
\caption{Parameter sensitivity of the $K^*\bar{K}^*$ (top) and $\eta\eta^\prime$ (bottom) two-body partial widths. All curvatures indicate the dependence of $K^*\bar{K}^*$ and $\eta\eta^\prime$ widths on the parameter $\Lambda_0$ with the $\Lambda_1$ and the ratio of $D$-wave component fixed. Solid/Dashed/DotDashed lines show the results with $5\%$/$4\%$/$3\%$ $D$-wave component and red/orange/blue lines show those with $\Lambda_1=1.5$/$2.0$/$2.5$~GeV. }\label{fig:kstarpw}
\end{figure}

The decay patterns of $1^{--}$ and $1^{-+}$ $K\bar K_1(1400)$ molecules for $\Lambda_0=1.3$~GeV and $\Lambda_1=2.5$~GeV are shown in Table~\ref{tab:widths}. Our results show that the three-body $K\bar{K}^*\pi$ is the largest decay channel for both $1^{--}$ and $1^{-+}$ $K\bar K_1$ molecules. And their dominant two-body decay channel is also the same, i.e., $K^*\bar{K}^*$. Moreover, the $1^{-+}$ exotic state also couples strongly to the $\eta\eta^\prime$ channel. It is quite similar with the properties of $D\bar{D}_1$ molecules where the $D^*\bar{D}^*$ and $\eta\eta_c$ channels contribute dominantly to the two-body decay widths~\cite{Dong:2019ofp}. 
In general, one can find that the molecular scenario always seems to support the strong couplings with the open-flavor channels. In addition, we find that the $1^{-+}$ molecule has also a strong coupling with $a_1 \pi$ channel, which is also reflected in a recent lattice simulation~\cite{Woss:2020ayi}. In that work, the exotic $1^{-+}$ hybird meson with negative $G$-parity decays dominantly into $b_1 \pi$ channel which translates to $a_1 \pi$ channel for the positive $G$-parity $1^{-+}$ state. The parameter sensitivity of partial widths of two dominant two-body decay channels for the $1^{-+}$ $K\bar K_1(1400)$ molecule, namely $K^*\bar{K}^*$ and $\eta\eta^\prime$, is presented in Fig.~\ref{fig:kstarpw}.

\section{Summary}
We have used the one boson exchange potential between the $K\bar K_1$, for both $J^{PC}=1^{--}$ and $J^{PC}=1^{-+}$ systems, to investigate if it is possible for them to form bound states. It turns out that with a momentum cutoff $\Lambda\approx 2.5$ GeV, the attractive force between $K\bar K_1(1400)$ with $J^{PC}=1^{-+}$ is strong enough to form a bound state with a binding energy of around 40 MeV, which can be related to the newly observed $\eta_1(1855)$. The C-parity partner of this molecule is predicted to exist with a binding energy around 10 $\sim$ 30 MeV. While for the $K_1(1270)$, the strong coupling between $K_1(1270)KV$ results in a repulsive force for $C=+$ and a much deeply attractive force for $C=-$, neither of which is expected to form a molecular state.

The decay properties of the two possible molecules of $K\bar K_1(1400)$ are investigated. For both the $1^{--}$ and $1^{-+}$ states, the three body channel $K\bar K^* \pi$ dominates due to the large decay width of $K_1(1400)\to K^*\pi$. The $K^*\bar K^*$ and $\eta\eta'$ channels have the largest contributions among the two body decays. The presented decay pattern of $1^{-+}$ $K\bar K_1(1400)$ molecule is compatible with BESIII's measurements.

In summary, it is shown that the newly observed exotic state $\eta_1(1855)$ can be explained as a $K\bar K_1(1400)$ molecule. This interpretation can be checked by searching for $\eta_1(1855)$ in $\omega\phi$ channel and its $C$-partner in $\pi\rho$, $\eta\omega$ and $K\bar K$ channels, in addition to $K^*\bar K^*$ channel.

\vspace{0.5cm}
We thank Feng-Kun Guo, Xiao-Yu Li, Jia-Jun Wu { and Mao-Jun Yan} for useful discussions. This work is supported by the NSFC and the Deutsche Forschungsgemeinschaft (DFG, German Research Foundation) through the funds provided to the Sino German Collaborative Research Center TRR110 Symmetries
and the Emergence of Structure in QCD (NSFC Grant No.12070131001, DFG Project-ID 196253076-TRR 110), by the
NSFC Grants No. 11835015 and No. 12047503, and by
the Chinese Academy of Sciences (CAS) under Grant No.XDB34030000.

\bibliography{ref}

\begin{thebibliography}{74}%
\makeatletter
\providecommand \@ifxundefined [1]{%
 \@ifx{#1\undefined}
}%
\providecommand \@ifnum [1]{%
 \ifnum #1\expandafter \@firstoftwo
 \else \expandafter \@secondoftwo
 \fi
}%
\providecommand \@ifx [1]{%
 \ifx #1\expandafter \@firstoftwo
 \else \expandafter \@secondoftwo
 \fi
}%
\providecommand \natexlab [1]{#1}%
\providecommand \enquote  [1]{``#1''}%
\providecommand \bibnamefont  [1]{#1}%
\providecommand \bibfnamefont [1]{#1}%
\providecommand \citenamefont [1]{#1}%
\providecommand \href@noop [0]{\@secondoftwo}%
\providecommand \href [0]{\begingroup \@sanitize@url \@href}%
\providecommand \@href[1]{\@@startlink{#1}\@@href}%
\providecommand \@@href[1]{\endgroup#1\@@endlink}%
\providecommand \@sanitize@url [0]{\catcode `\\12\catcode `\$12\catcode
  `\&12\catcode `\#12\catcode `\^12\catcode `\_12\catcode `\%12\relax}%
\providecommand \@@startlink[1]{}%
\providecommand \@@endlink[0]{}%
\providecommand \url  [0]{\begingroup\@sanitize@url \@url }%
\providecommand \@url [1]{\endgroup\@href {#1}{\urlprefix }}%
\providecommand \urlprefix  [0]{URL }%
\providecommand \Eprint [0]{\href }%
\providecommand \doibase [0]{http://dx.doi.org/}%
\providecommand \selectlanguage [0]{\@gobble}%
\providecommand \bibinfo  [0]{\@secondoftwo}%
\providecommand \bibfield  [0]{\@secondoftwo}%
\providecommand \translation [1]{[#1]}%
\providecommand \BibitemOpen [0]{}%
\providecommand \bibitemStop [0]{}%
\providecommand \bibitemNoStop [0]{.\EOS\space}%
\providecommand \EOS [0]{\spacefactor3000\relax}%
\providecommand \BibitemShut  [1]{\csname bibitem#1\endcsname}%
\let\auto@bib@innerbib\@empty
\bibitem [{\citenamefont {Gell-Mann}(1964)}]{GellMann:1964nj}%
  \BibitemOpen
  \bibfield  {author} {\bibinfo {author} {\bibfnamefont {M.}~\bibnamefont
  {Gell-Mann}},\ }\href {\doibase 10.1016/S0031-9163(64)92001-3} {\bibfield
  {journal} {\bibinfo  {journal} {Phys. Lett.}\ }\textbf {\bibinfo {volume}
  {8}},\ \bibinfo {pages} {214} (\bibinfo {year} {1964})}\BibitemShut {NoStop}%
\bibitem [{\citenamefont {Zweig}(1964)}]{Zweig:1964jf}%
  \BibitemOpen
  \bibfield  {author} {\bibinfo {author} {\bibfnamefont {G.}~\bibnamefont
  {Zweig}},\ }\enquote {\bibinfo {title} {{An SU(3) model for strong
  interaction symmetry and its breaking. Version 2}},}\ in\ \href@noop {}
  {\emph {\bibinfo {booktitle} {{DEVELOPMENTS IN THE QUARK THEORY OF HADRONS.
  VOL. 1. 1964 - 1978}}}},\ \bibinfo {editor} {edited by\ \bibinfo {editor}
  {\bibfnamefont {D.~B.}\ \bibnamefont {Lichtenberg}}\ and\ \bibinfo {editor}
  {\bibfnamefont {S.~P.}\ \bibnamefont {Rosen}}}\ (\bibinfo {year} {1964})\
  pp.\ \bibinfo {pages} {22--101}\BibitemShut {NoStop}%
\bibitem [{\citenamefont {Klempt}\ and\ \citenamefont
  {Zaitsev}(2007)}]{Klempt:2007cp}%
  \BibitemOpen
  \bibfield  {author} {\bibinfo {author} {\bibfnamefont {E.}~\bibnamefont
  {Klempt}}\ and\ \bibinfo {author} {\bibfnamefont {A.}~\bibnamefont
  {Zaitsev}},\ }\href {\doibase 10.1016/j.physrep.2007.07.006} {\bibfield
  {journal} {\bibinfo  {journal} {Phys. Rept.}\ }\textbf {\bibinfo {volume}
  {454}},\ \bibinfo {pages} {1} (\bibinfo {year} {2007})},\ \Eprint
  {http://arxiv.org/abs/0708.4016} {arXiv:0708.4016 [hep-ph]} \BibitemShut
  {NoStop}%
\bibitem [{\citenamefont {Dalitz}\ and\ \citenamefont
  {Tuan}(1959)}]{Dalitz:1959dn}%
  \BibitemOpen
  \bibfield  {author} {\bibinfo {author} {\bibfnamefont {R.~H.}\ \bibnamefont
  {Dalitz}}\ and\ \bibinfo {author} {\bibfnamefont {S.~F.}\ \bibnamefont
  {Tuan}},\ }\href {\doibase 10.1103/PhysRevLett.2.425} {\bibfield  {journal}
  {\bibinfo  {journal} {Phys. Rev. Lett.}\ }\textbf {\bibinfo {volume} {2}},\
  \bibinfo {pages} {425} (\bibinfo {year} {1959})}\BibitemShut {NoStop}%
\bibitem [{\citenamefont {Dalitz}\ and\ \citenamefont
  {Tuan}(1960)}]{Dalitz:1960du}%
  \BibitemOpen
  \bibfield  {author} {\bibinfo {author} {\bibfnamefont {R.~H.}\ \bibnamefont
  {Dalitz}}\ and\ \bibinfo {author} {\bibfnamefont {S.~F.}\ \bibnamefont
  {Tuan}},\ }\href {\doibase 10.1016/0003-4916(60)90001-4} {\bibfield
  {journal} {\bibinfo  {journal} {Annals Phys.}\ }\textbf {\bibinfo {volume}
  {10}},\ \bibinfo {pages} {307} (\bibinfo {year} {1960})}\BibitemShut
  {NoStop}%
\bibitem [{\citenamefont {Alston}\ \emph {et~al.}(1961)\citenamefont {Alston},
  \citenamefont {Alvarez}, \citenamefont {Eberhard}, \citenamefont {Good},
  \citenamefont {Graziano}, \citenamefont {Ticho},\ and\ \citenamefont
  {Wojcicki}}]{Alston:1961zzd}%
  \BibitemOpen
  \bibfield  {author} {\bibinfo {author} {\bibfnamefont {M.~H.}\ \bibnamefont
  {Alston}}, \bibinfo {author} {\bibfnamefont {L.~W.}\ \bibnamefont {Alvarez}},
  \bibinfo {author} {\bibfnamefont {P.}~\bibnamefont {Eberhard}}, \bibinfo
  {author} {\bibfnamefont {M.~L.}\ \bibnamefont {Good}}, \bibinfo {author}
  {\bibfnamefont {W.}~\bibnamefont {Graziano}}, \bibinfo {author}
  {\bibfnamefont {H.~K.}\ \bibnamefont {Ticho}}, \ and\ \bibinfo {author}
  {\bibfnamefont {S.~G.}\ \bibnamefont {Wojcicki}},\ }\href {\doibase
  10.1103/PhysRevLett.6.698} {\bibfield  {journal} {\bibinfo  {journal} {Phys.
  Rev. Lett.}\ }\textbf {\bibinfo {volume} {6}},\ \bibinfo {pages} {698}
  (\bibinfo {year} {1961})}\BibitemShut {NoStop}%
\bibitem [{\citenamefont {Hyodo}\ and\ \citenamefont
  {Niiyama}(2021)}]{Hyodo:2020czb}%
  \BibitemOpen
  \bibfield  {author} {\bibinfo {author} {\bibfnamefont {T.}~\bibnamefont
  {Hyodo}}\ and\ \bibinfo {author} {\bibfnamefont {M.}~\bibnamefont
  {Niiyama}},\ }\href {\doibase 10.1016/j.ppnp.2021.103868} {\bibfield
  {journal} {\bibinfo  {journal} {Prog. Part. Nucl. Phys.}\ }\textbf {\bibinfo
  {volume} {120}},\ \bibinfo {pages} {103868} (\bibinfo {year} {2021})},\
  \Eprint {http://arxiv.org/abs/2010.07592} {arXiv:2010.07592 [hep-ph]}
  \BibitemShut {NoStop}%
\bibitem [{\citenamefont {Mai}(2021)}]{Mai:2020ltx}%
  \BibitemOpen
  \bibfield  {author} {\bibinfo {author} {\bibfnamefont {M.}~\bibnamefont
  {Mai}},\ }\href {\doibase 10.1140/epjs/s11734-021-00144-7} {\bibfield
  {journal} {\bibinfo  {journal} {Eur. Phys. J. ST}\ }\textbf {\bibinfo
  {volume} {230}},\ \bibinfo {pages} {1593} (\bibinfo {year} {2021})},\ \Eprint
  {http://arxiv.org/abs/2010.00056} {arXiv:2010.00056 [nucl-th]} \BibitemShut
  {NoStop}%
\bibitem [{\citenamefont {Choi}\ \emph {et~al.}(2003)\citenamefont {Choi} \emph
  {et~al.}}]{Choi:2003ue}%
  \BibitemOpen
  \bibfield  {author} {\bibinfo {author} {\bibfnamefont {S.~K.}\ \bibnamefont
  {Choi}} \emph {et~al.} (\bibinfo {collaboration} {Belle}),\ }\href {\doibase
  10.1103/PhysRevLett.91.262001} {\bibfield  {journal} {\bibinfo  {journal}
  {Phys. Rev. Lett.}\ }\textbf {\bibinfo {volume} {91}},\ \bibinfo {pages}
  {262001} (\bibinfo {year} {2003})},\ \Eprint
  {http://arxiv.org/abs/hep-ex/0309032} {arXiv:hep-ex/0309032} \BibitemShut
  {NoStop}%
\bibitem [{\citenamefont {Chen}\ \emph {et~al.}(2016)\citenamefont {Chen},
  \citenamefont {Chen}, \citenamefont {Liu},\ and\ \citenamefont
  {Zhu}}]{Chen:2016qju}%
  \BibitemOpen
  \bibfield  {author} {\bibinfo {author} {\bibfnamefont {H.-X.}\ \bibnamefont
  {Chen}}, \bibinfo {author} {\bibfnamefont {W.}~\bibnamefont {Chen}}, \bibinfo
  {author} {\bibfnamefont {X.}~\bibnamefont {Liu}}, \ and\ \bibinfo {author}
  {\bibfnamefont {S.-L.}\ \bibnamefont {Zhu}},\ }\href {\doibase
  10.1016/j.physrep.2016.05.004} {\bibfield  {journal} {\bibinfo  {journal}
  {Phys. Rept.}\ }\textbf {\bibinfo {volume} {639}},\ \bibinfo {pages} {1}
  (\bibinfo {year} {2016})},\ \Eprint {http://arxiv.org/abs/1601.02092}
  {arXiv:1601.02092 [hep-ph]} \BibitemShut {NoStop}%
\bibitem [{\citenamefont {Hosaka}\ \emph {et~al.}(2016)\citenamefont {Hosaka},
  \citenamefont {Iijima}, \citenamefont {Miyabayashi}, \citenamefont {Sakai},\
  and\ \citenamefont {Yasui}}]{Hosaka:2016pey}%
  \BibitemOpen
  \bibfield  {author} {\bibinfo {author} {\bibfnamefont {A.}~\bibnamefont
  {Hosaka}}, \bibinfo {author} {\bibfnamefont {T.}~\bibnamefont {Iijima}},
  \bibinfo {author} {\bibfnamefont {K.}~\bibnamefont {Miyabayashi}}, \bibinfo
  {author} {\bibfnamefont {Y.}~\bibnamefont {Sakai}}, \ and\ \bibinfo {author}
  {\bibfnamefont {S.}~\bibnamefont {Yasui}},\ }\href {\doibase
  10.1093/ptep/ptw045} {\bibfield  {journal} {\bibinfo  {journal} {PTEP}\
  }\textbf {\bibinfo {volume} {2016}},\ \bibinfo {pages} {062C01} (\bibinfo
  {year} {2016})},\ \Eprint {http://arxiv.org/abs/1603.09229} {arXiv:1603.09229
  [hep-ph]} \BibitemShut {NoStop}%
\bibitem [{\citenamefont {Richard}(2016)}]{Richard:2016eis}%
  \BibitemOpen
  \bibfield  {author} {\bibinfo {author} {\bibfnamefont {J.-M.}\ \bibnamefont
  {Richard}},\ }\href {\doibase 10.1007/s00601-016-1159-0} {\bibfield
  {journal} {\bibinfo  {journal} {Few Body Syst.}\ }\textbf {\bibinfo {volume}
  {57}},\ \bibinfo {pages} {1185} (\bibinfo {year} {2016})},\ \Eprint
  {http://arxiv.org/abs/1606.08593} {arXiv:1606.08593 [hep-ph]} \BibitemShut
  {NoStop}%
\bibitem [{\citenamefont {Lebed}\ \emph {et~al.}(2017)\citenamefont {Lebed},
  \citenamefont {Mitchell},\ and\ \citenamefont {Swanson}}]{Lebed:2016hpi}%
  \BibitemOpen
  \bibfield  {author} {\bibinfo {author} {\bibfnamefont {R.~F.}\ \bibnamefont
  {Lebed}}, \bibinfo {author} {\bibfnamefont {R.~E.}\ \bibnamefont {Mitchell}},
  \ and\ \bibinfo {author} {\bibfnamefont {E.~S.}\ \bibnamefont {Swanson}},\
  }\href {\doibase 10.1016/j.ppnp.2016.11.003} {\bibfield  {journal} {\bibinfo
  {journal} {Prog. Part. Nucl. Phys.}\ }\textbf {\bibinfo {volume} {93}},\
  \bibinfo {pages} {143} (\bibinfo {year} {2017})},\ \Eprint
  {http://arxiv.org/abs/1610.04528} {arXiv:1610.04528 [hep-ph]} \BibitemShut
  {NoStop}%
\bibitem [{\citenamefont {Esposito}\ \emph {et~al.}(2017)\citenamefont
  {Esposito}, \citenamefont {Pilloni},\ and\ \citenamefont
  {Polosa}}]{Esposito:2016noz}%
  \BibitemOpen
  \bibfield  {author} {\bibinfo {author} {\bibfnamefont {A.}~\bibnamefont
  {Esposito}}, \bibinfo {author} {\bibfnamefont {A.}~\bibnamefont {Pilloni}}, \
  and\ \bibinfo {author} {\bibfnamefont {A.~D.}\ \bibnamefont {Polosa}},\
  }\href {\doibase 10.1016/j.physrep.2016.11.002} {\bibfield  {journal}
  {\bibinfo  {journal} {Phys. Rept.}\ }\textbf {\bibinfo {volume} {668}},\
  \bibinfo {pages} {1} (\bibinfo {year} {2017})},\ \Eprint
  {http://arxiv.org/abs/1611.07920} {arXiv:1611.07920 [hep-ph]} \BibitemShut
  {NoStop}%
\bibitem [{\citenamefont {Guo}\ \emph {et~al.}(2018)\citenamefont {Guo},
  \citenamefont {Hanhart}, \citenamefont {Mei\ss{}ner}, \citenamefont {Wang},
  \citenamefont {Zhao},\ and\ \citenamefont {Zou}}]{Guo:2017jvc}%
  \BibitemOpen
  \bibfield  {author} {\bibinfo {author} {\bibfnamefont {F.-K.}\ \bibnamefont
  {Guo}}, \bibinfo {author} {\bibfnamefont {C.}~\bibnamefont {Hanhart}},
  \bibinfo {author} {\bibfnamefont {U.-G.}\ \bibnamefont {Mei\ss{}ner}},
  \bibinfo {author} {\bibfnamefont {Q.}~\bibnamefont {Wang}}, \bibinfo {author}
  {\bibfnamefont {Q.}~\bibnamefont {Zhao}}, \ and\ \bibinfo {author}
  {\bibfnamefont {B.-S.}\ \bibnamefont {Zou}},\ }\href {\doibase
  10.1103/RevModPhys.90.015004} {\bibfield  {journal} {\bibinfo  {journal}
  {Rev. Mod. Phys.}\ }\textbf {\bibinfo {volume} {90}},\ \bibinfo {pages}
  {015004} (\bibinfo {year} {2018})},\ \Eprint
  {http://arxiv.org/abs/1705.00141} {arXiv:1705.00141 [hep-ph]} \BibitemShut
  {NoStop}%
\bibitem [{\citenamefont {Ali}\ \emph {et~al.}(2017)\citenamefont {Ali},
  \citenamefont {Lange},\ and\ \citenamefont {Stone}}]{Ali:2017jda}%
  \BibitemOpen
  \bibfield  {author} {\bibinfo {author} {\bibfnamefont {A.}~\bibnamefont
  {Ali}}, \bibinfo {author} {\bibfnamefont {J.~S.}\ \bibnamefont {Lange}}, \
  and\ \bibinfo {author} {\bibfnamefont {S.}~\bibnamefont {Stone}},\ }\href
  {\doibase 10.1016/j.ppnp.2017.08.003} {\bibfield  {journal} {\bibinfo
  {journal} {Prog. Part. Nucl. Phys.}\ }\textbf {\bibinfo {volume} {97}},\
  \bibinfo {pages} {123} (\bibinfo {year} {2017})},\ \Eprint
  {http://arxiv.org/abs/1706.00610} {arXiv:1706.00610 [hep-ph]} \BibitemShut
  {NoStop}%
\bibitem [{\citenamefont {Olsen}\ \emph {et~al.}(2018)\citenamefont {Olsen},
  \citenamefont {Skwarnicki},\ and\ \citenamefont {Zieminska}}]{Olsen:2017bmm}%
  \BibitemOpen
  \bibfield  {author} {\bibinfo {author} {\bibfnamefont {S.~L.}\ \bibnamefont
  {Olsen}}, \bibinfo {author} {\bibfnamefont {T.}~\bibnamefont {Skwarnicki}}, \
  and\ \bibinfo {author} {\bibfnamefont {D.}~\bibnamefont {Zieminska}},\ }\href
  {\doibase 10.1103/RevModPhys.90.015003} {\bibfield  {journal} {\bibinfo
  {journal} {Rev. Mod. Phys.}\ }\textbf {\bibinfo {volume} {90}},\ \bibinfo
  {pages} {015003} (\bibinfo {year} {2018})},\ \Eprint
  {http://arxiv.org/abs/1708.04012} {arXiv:1708.04012 [hep-ph]} \BibitemShut
  {NoStop}%
\bibitem [{\citenamefont {Altmannshofer}\ \emph {et~al.}(2019)\citenamefont
  {Altmannshofer} \emph {et~al.}}]{Kou:2018nap}%
  \BibitemOpen
  \bibfield  {author} {\bibinfo {author} {\bibfnamefont {W.}~\bibnamefont
  {Altmannshofer}} \emph {et~al.} (\bibinfo {collaboration} {Belle-II}),\
  }\href {\doibase 10.1093/ptep/ptz106} {\bibfield  {journal} {\bibinfo
  {journal} {PTEP}\ }\textbf {\bibinfo {volume} {2019}},\ \bibinfo {pages}
  {123C01} (\bibinfo {year} {2019})},\ \bibinfo {note} {[Erratum: PTEP 2020,
  029201 (2020)]},\ \Eprint {http://arxiv.org/abs/1808.10567} {arXiv:1808.10567
  [hep-ex]} \BibitemShut {NoStop}%
\bibitem [{\citenamefont {Kalashnikova}\ and\ \citenamefont
  {Nefediev}(2019)}]{Kalashnikova:2018vkv}%
  \BibitemOpen
  \bibfield  {author} {\bibinfo {author} {\bibfnamefont {Y.~S.}\ \bibnamefont
  {Kalashnikova}}\ and\ \bibinfo {author} {\bibfnamefont {A.~V.}\ \bibnamefont
  {Nefediev}},\ }\href {\doibase 10.3367/UFNe.2018.08.038411} {\bibfield
  {journal} {\bibinfo  {journal} {Phys. Usp.}\ }\textbf {\bibinfo {volume}
  {62}},\ \bibinfo {pages} {568} (\bibinfo {year} {2019})},\ \Eprint
  {http://arxiv.org/abs/1811.01324} {arXiv:1811.01324 [hep-ph]} \BibitemShut
  {NoStop}%
\bibitem [{\citenamefont {Cerri}\ \emph {et~al.}(2019)\citenamefont {Cerri}
  \emph {et~al.}}]{Cerri:2018ypt}%
  \BibitemOpen
  \bibfield  {author} {\bibinfo {author} {\bibfnamefont {A.}~\bibnamefont
  {Cerri}} \emph {et~al.},\ }\href {\doibase 10.23731/CYRM-2019-007.867}
  {\bibfield  {journal} {\bibinfo  {journal} {CERN Yellow Rep. Monogr.}\
  }\textbf {\bibinfo {volume} {7}},\ \bibinfo {pages} {867} (\bibinfo {year}
  {2019})},\ \Eprint {http://arxiv.org/abs/1812.07638} {arXiv:1812.07638
  [hep-ph]} \BibitemShut {NoStop}%
\bibitem [{\citenamefont {Liu}\ \emph {et~al.}(2019)\citenamefont {Liu},
  \citenamefont {Chen}, \citenamefont {Chen}, \citenamefont {Liu},\ and\
  \citenamefont {Zhu}}]{Liu:2019zoy}%
  \BibitemOpen
  \bibfield  {author} {\bibinfo {author} {\bibfnamefont {Y.-R.}\ \bibnamefont
  {Liu}}, \bibinfo {author} {\bibfnamefont {H.-X.}\ \bibnamefont {Chen}},
  \bibinfo {author} {\bibfnamefont {W.}~\bibnamefont {Chen}}, \bibinfo {author}
  {\bibfnamefont {X.}~\bibnamefont {Liu}}, \ and\ \bibinfo {author}
  {\bibfnamefont {S.-L.}\ \bibnamefont {Zhu}},\ }\href {\doibase
  10.1016/j.ppnp.2019.04.003} {\bibfield  {journal} {\bibinfo  {journal} {Prog.
  Part. Nucl. Phys.}\ }\textbf {\bibinfo {volume} {107}},\ \bibinfo {pages}
  {237} (\bibinfo {year} {2019})},\ \Eprint {http://arxiv.org/abs/1903.11976}
  {arXiv:1903.11976 [hep-ph]} \BibitemShut {NoStop}%
\bibitem [{\citenamefont {Brambilla}\ \emph {et~al.}(2020)\citenamefont
  {Brambilla}, \citenamefont {Eidelman}, \citenamefont {Hanhart}, \citenamefont
  {Nefediev}, \citenamefont {Shen}, \citenamefont {Thomas}, \citenamefont
  {Vairo},\ and\ \citenamefont {Yuan}}]{Brambilla:2019esw}%
  \BibitemOpen
  \bibfield  {author} {\bibinfo {author} {\bibfnamefont {N.}~\bibnamefont
  {Brambilla}}, \bibinfo {author} {\bibfnamefont {S.}~\bibnamefont {Eidelman}},
  \bibinfo {author} {\bibfnamefont {C.}~\bibnamefont {Hanhart}}, \bibinfo
  {author} {\bibfnamefont {A.}~\bibnamefont {Nefediev}}, \bibinfo {author}
  {\bibfnamefont {C.-P.}\ \bibnamefont {Shen}}, \bibinfo {author}
  {\bibfnamefont {C.~E.}\ \bibnamefont {Thomas}}, \bibinfo {author}
  {\bibfnamefont {A.}~\bibnamefont {Vairo}}, \ and\ \bibinfo {author}
  {\bibfnamefont {C.-Z.}\ \bibnamefont {Yuan}},\ }\href {\doibase
  10.1016/j.physrep.2020.05.001} {\bibfield  {journal} {\bibinfo  {journal}
  {Phys. Rept.}\ }\textbf {\bibinfo {volume} {873}},\ \bibinfo {pages} {1}
  (\bibinfo {year} {2020})},\ \Eprint {http://arxiv.org/abs/1907.07583}
  {arXiv:1907.07583 [hep-ex]} \BibitemShut {NoStop}%
\bibitem [{\citenamefont {Guo}\ \emph {et~al.}(2020)\citenamefont {Guo},
  \citenamefont {Liu},\ and\ \citenamefont {Sakai}}]{Guo:2019twa}%
  \BibitemOpen
  \bibfield  {author} {\bibinfo {author} {\bibfnamefont {F.-K.}\ \bibnamefont
  {Guo}}, \bibinfo {author} {\bibfnamefont {X.-H.}\ \bibnamefont {Liu}}, \ and\
  \bibinfo {author} {\bibfnamefont {S.}~\bibnamefont {Sakai}},\ }\href
  {\doibase 10.1016/j.ppnp.2020.103757} {\bibfield  {journal} {\bibinfo
  {journal} {Prog. Part. Nucl. Phys.}\ }\textbf {\bibinfo {volume} {112}},\
  \bibinfo {pages} {103757} (\bibinfo {year} {2020})},\ \Eprint
  {http://arxiv.org/abs/1912.07030} {arXiv:1912.07030 [hep-ph]} \BibitemShut
  {NoStop}%
\bibitem [{\citenamefont {Yang}\ \emph {et~al.}(2020)\citenamefont {Yang},
  \citenamefont {Ping},\ and\ \citenamefont {Segovia}}]{Yang:2020atz}%
  \BibitemOpen
  \bibfield  {author} {\bibinfo {author} {\bibfnamefont {G.}~\bibnamefont
  {Yang}}, \bibinfo {author} {\bibfnamefont {J.}~\bibnamefont {Ping}}, \ and\
  \bibinfo {author} {\bibfnamefont {J.}~\bibnamefont {Segovia}},\ }\href
  {\doibase 10.3390/sym12111869} {\bibfield  {journal} {\bibinfo  {journal}
  {Symmetry}\ }\textbf {\bibinfo {volume} {12}},\ \bibinfo {pages} {1869}
  (\bibinfo {year} {2020})},\ \Eprint {http://arxiv.org/abs/2009.00238}
  {arXiv:2009.00238 [hep-ph]} \BibitemShut {NoStop}%
\bibitem [{\citenamefont {Ortega}\ and\ \citenamefont
  {Entem}(2021)}]{Ortega:2020tng}%
  \BibitemOpen
  \bibfield  {author} {\bibinfo {author} {\bibfnamefont {P.~G.}\ \bibnamefont
  {Ortega}}\ and\ \bibinfo {author} {\bibfnamefont {D.~R.}\ \bibnamefont
  {Entem}},\ }\href {\doibase 10.3390/sym13020279} {\bibfield  {journal}
  {\bibinfo  {journal} {Symmetry}\ }\textbf {\bibinfo {volume} {13}},\ \bibinfo
  {pages} {279} (\bibinfo {year} {2021})},\ \Eprint
  {http://arxiv.org/abs/2012.10105} {arXiv:2012.10105 [hep-ph]} \BibitemShut
  {NoStop}%
\bibitem [{\citenamefont {Dong}\ \emph
  {et~al.}(2021{\natexlab{a}})\citenamefont {Dong}, \citenamefont {Guo},\ and\
  \citenamefont {Zou}}]{Dong:2021juy}%
  \BibitemOpen
  \bibfield  {author} {\bibinfo {author} {\bibfnamefont {X.-K.}\ \bibnamefont
  {Dong}}, \bibinfo {author} {\bibfnamefont {F.-K.}\ \bibnamefont {Guo}}, \
  and\ \bibinfo {author} {\bibfnamefont {B.-S.}\ \bibnamefont {Zou}},\ }\href
  {\doibase 10.13725/j.cnki.pip.2021.02.001} {\bibfield  {journal} {\bibinfo
  {journal} {Progr. Phys.}\ }\textbf {\bibinfo {volume} {41}},\ \bibinfo
  {pages} {65} (\bibinfo {year} {2021}{\natexlab{a}})},\ \Eprint
  {http://arxiv.org/abs/2101.01021} {arXiv:2101.01021 [hep-ph]} \BibitemShut
  {NoStop}%
\bibitem [{\citenamefont {Dong}\ \emph
  {et~al.}(2021{\natexlab{b}})\citenamefont {Dong}, \citenamefont {Guo},\ and\
  \citenamefont {Zou}}]{Dong:2021bvy}%
  \BibitemOpen
  \bibfield  {author} {\bibinfo {author} {\bibfnamefont {X.-K.}\ \bibnamefont
  {Dong}}, \bibinfo {author} {\bibfnamefont {F.-K.}\ \bibnamefont {Guo}}, \
  and\ \bibinfo {author} {\bibfnamefont {B.-S.}\ \bibnamefont {Zou}},\ }\href
  {\doibase 10.1088/1572-9494/ac27a2} {\bibfield  {journal} {\bibinfo
  {journal} {Commun. Theor. Phys.}\ }\textbf {\bibinfo {volume} {73}},\
  \bibinfo {pages} {125201} (\bibinfo {year} {2021}{\natexlab{b}})},\ \Eprint
  {http://arxiv.org/abs/2108.02673} {arXiv:2108.02673 [hep-ph]} \BibitemShut
  {NoStop}%
\bibitem [{\citenamefont {Godfrey}\ and\ \citenamefont
  {Isgur}(1985)}]{Godfrey:1985xj}%
  \BibitemOpen
  \bibfield  {author} {\bibinfo {author} {\bibfnamefont {S.}~\bibnamefont
  {Godfrey}}\ and\ \bibinfo {author} {\bibfnamefont {N.}~\bibnamefont
  {Isgur}},\ }\href {\doibase 10.1103/PhysRevD.32.189} {\bibfield  {journal}
  {\bibinfo  {journal} {Phys. Rev. D}\ }\textbf {\bibinfo {volume} {32}},\
  \bibinfo {pages} {189} (\bibinfo {year} {1985})}\BibitemShut {NoStop}%
\bibitem [{\citenamefont {Capstick}\ and\ \citenamefont
  {Isgur}(1985)}]{Capstick:1985xss}%
  \BibitemOpen
  \bibfield  {author} {\bibinfo {author} {\bibfnamefont {S.}~\bibnamefont
  {Capstick}}\ and\ \bibinfo {author} {\bibfnamefont {N.}~\bibnamefont
  {Isgur}},\ }\href {\doibase 10.1063/1.35361} {\bibfield  {journal} {\bibinfo
  {journal} {AIP Conf. Proc.}\ }\textbf {\bibinfo {volume} {132}},\ \bibinfo
  {pages} {267} (\bibinfo {year} {1985})}\BibitemShut {NoStop}%
\bibitem [{\citenamefont {Aubert}\ \emph {et~al.}(2005)\citenamefont {Aubert}
  \emph {et~al.}}]{BaBar:2005hhc}%
  \BibitemOpen
  \bibfield  {author} {\bibinfo {author} {\bibfnamefont {B.}~\bibnamefont
  {Aubert}} \emph {et~al.} (\bibinfo {collaboration} {BaBar}),\ }\href
  {\doibase 10.1103/PhysRevLett.95.142001} {\bibfield  {journal} {\bibinfo
  {journal} {Phys. Rev. Lett.}\ }\textbf {\bibinfo {volume} {95}},\ \bibinfo
  {pages} {142001} (\bibinfo {year} {2005})},\ \Eprint
  {http://arxiv.org/abs/hep-ex/0506081} {arXiv:hep-ex/0506081} \BibitemShut
  {NoStop}%
\bibitem [{\citenamefont {Ablikim}\ \emph {et~al.}(2013)\citenamefont {Ablikim}
  \emph {et~al.}}]{Ablikim:2013mio}%
  \BibitemOpen
  \bibfield  {author} {\bibinfo {author} {\bibfnamefont {M.}~\bibnamefont
  {Ablikim}} \emph {et~al.} (\bibinfo {collaboration} {BESIII}),\ }\href
  {\doibase 10.1103/PhysRevLett.110.252001} {\bibfield  {journal} {\bibinfo
  {journal} {Phys. Rev. Lett.}\ }\textbf {\bibinfo {volume} {110}},\ \bibinfo
  {pages} {252001} (\bibinfo {year} {2013})},\ \Eprint
  {http://arxiv.org/abs/1303.5949} {arXiv:1303.5949 [hep-ex]} \BibitemShut
  {NoStop}%
\bibitem [{\citenamefont {Liu}\ \emph {et~al.}(2013)\citenamefont {Liu} \emph
  {et~al.}}]{Liu:2013dau}%
  \BibitemOpen
  \bibfield  {author} {\bibinfo {author} {\bibfnamefont {Z.~Q.}\ \bibnamefont
  {Liu}} \emph {et~al.} (\bibinfo {collaboration} {Belle}),\ }\href {\doibase
  10.1103/PhysRevLett.110.252002} {\bibfield  {journal} {\bibinfo  {journal}
  {Phys. Rev. Lett.}\ }\textbf {\bibinfo {volume} {110}},\ \bibinfo {pages}
  {252002} (\bibinfo {year} {2013})},\ \bibinfo {note} {[Erratum:
  Phys.Rev.Lett. 111, 019901 (2013)]},\ \Eprint
  {http://arxiv.org/abs/1304.0121} {arXiv:1304.0121 [hep-ex]} \BibitemShut
  {NoStop}%
\bibitem [{\citenamefont {Ablikim}\ \emph {et~al.}(2021)\citenamefont {Ablikim}
  \emph {et~al.}}]{BESIII:2020qkh}%
  \BibitemOpen
  \bibfield  {author} {\bibinfo {author} {\bibfnamefont {M.}~\bibnamefont
  {Ablikim}} \emph {et~al.} (\bibinfo {collaboration} {BESIII}),\ }\href
  {\doibase 10.1103/PhysRevLett.126.102001} {\bibfield  {journal} {\bibinfo
  {journal} {Phys. Rev. Lett.}\ }\textbf {\bibinfo {volume} {126}},\ \bibinfo
  {pages} {102001} (\bibinfo {year} {2021})},\ \Eprint
  {http://arxiv.org/abs/2011.07855} {arXiv:2011.07855 [hep-ex]} \BibitemShut
  {NoStop}%
\bibitem [{\citenamefont {Aaij}\ \emph {et~al.}(2015)\citenamefont {Aaij} \emph
  {et~al.}}]{Aaij:2015tga}%
  \BibitemOpen
  \bibfield  {author} {\bibinfo {author} {\bibfnamefont {R.}~\bibnamefont
  {Aaij}} \emph {et~al.} (\bibinfo {collaboration} {LHCb}),\ }\href {\doibase
  10.1103/PhysRevLett.115.072001} {\bibfield  {journal} {\bibinfo  {journal}
  {Phys. Rev. Lett.}\ }\textbf {\bibinfo {volume} {115}},\ \bibinfo {pages}
  {072001} (\bibinfo {year} {2015})},\ \Eprint
  {http://arxiv.org/abs/1507.03414} {arXiv:1507.03414 [hep-ex]} \BibitemShut
  {NoStop}%
\bibitem [{\citenamefont {Aaij}\ \emph {et~al.}(2019)\citenamefont {Aaij} \emph
  {et~al.}}]{Aaij:2019vzc}%
  \BibitemOpen
  \bibfield  {author} {\bibinfo {author} {\bibfnamefont {R.}~\bibnamefont
  {Aaij}} \emph {et~al.} (\bibinfo {collaboration} {LHCb}),\ }\href {\doibase
  10.1103/PhysRevLett.122.222001} {\bibfield  {journal} {\bibinfo  {journal}
  {Phys. Rev. Lett.}\ }\textbf {\bibinfo {volume} {122}},\ \bibinfo {pages}
  {222001} (\bibinfo {year} {2019})},\ \Eprint
  {http://arxiv.org/abs/1904.03947} {arXiv:1904.03947 [hep-ex]} \BibitemShut
  {NoStop}%
\bibitem [{\citenamefont {Aaij}\ \emph
  {et~al.}(2021{\natexlab{a}})\citenamefont {Aaij} \emph
  {et~al.}}]{LHCb:2020jpq}%
  \BibitemOpen
  \bibfield  {author} {\bibinfo {author} {\bibfnamefont {R.}~\bibnamefont
  {Aaij}} \emph {et~al.} (\bibinfo {collaboration} {LHCb}),\ }\href {\doibase
  10.1016/j.scib.2021.02.030} {\bibfield  {journal} {\bibinfo  {journal} {Sci.
  Bull.}\ }\textbf {\bibinfo {volume} {66}},\ \bibinfo {pages} {1391} (\bibinfo
  {year} {2021}{\natexlab{a}})},\ \Eprint {http://arxiv.org/abs/2012.10380}
  {arXiv:2012.10380 [hep-ex]} \BibitemShut {NoStop}%
\bibitem [{\citenamefont {Aaij}\ \emph
  {et~al.}(2021{\natexlab{b}})\citenamefont {Aaij} \emph
  {et~al.}}]{LHCb:2021vvq}%
  \BibitemOpen
  \bibfield  {author} {\bibinfo {author} {\bibfnamefont {R.}~\bibnamefont
  {Aaij}} \emph {et~al.} (\bibinfo {collaboration} {LHCb}),\ }\href@noop {} {\
  (\bibinfo {year} {2021}{\natexlab{b}})},\ \Eprint
  {http://arxiv.org/abs/2109.01038} {arXiv:2109.01038 [hep-ex]} \BibitemShut
  {NoStop}%
\bibitem [{\citenamefont {Aaij}\ \emph
  {et~al.}(2021{\natexlab{c}})\citenamefont {Aaij} \emph
  {et~al.}}]{LHCb:2021auc}%
  \BibitemOpen
  \bibfield  {author} {\bibinfo {author} {\bibfnamefont {R.}~\bibnamefont
  {Aaij}} \emph {et~al.} (\bibinfo {collaboration} {LHCb}),\ }\href@noop {} {\
  (\bibinfo {year} {2021}{\natexlab{c}})},\ \Eprint
  {http://arxiv.org/abs/2109.01056} {arXiv:2109.01056 [hep-ex]} \BibitemShut
  {NoStop}%
\bibitem [{\citenamefont {Zyla}\ \emph {et~al.}(2020)\citenamefont {Zyla} \emph
  {et~al.}}]{ParticleDataGroup:2020ssz}%
  \BibitemOpen
  \bibfield  {author} {\bibinfo {author} {\bibfnamefont {P.~A.}\ \bibnamefont
  {Zyla}} \emph {et~al.} (\bibinfo {collaboration} {Particle Data Group}),\
  }\href {\doibase 10.1093/ptep/ptaa104} {\bibfield  {journal} {\bibinfo
  {journal} {PTEP}\ }\textbf {\bibinfo {volume} {2020}},\ \bibinfo {pages}
  {083C01} (\bibinfo {year} {2020})}\BibitemShut {NoStop}%
\bibitem [{\citenamefont {Kuhn}\ \emph {et~al.}(2004)\citenamefont {Kuhn} \emph
  {et~al.}}]{Kuhn:2004en}%
  \BibitemOpen
  \bibfield  {author} {\bibinfo {author} {\bibfnamefont {J.}~\bibnamefont
  {Kuhn}} \emph {et~al.} (\bibinfo {collaboration} {E852}),\ }\href {\doibase
  10.1016/j.physletb.2004.05.032} {\bibfield  {journal} {\bibinfo  {journal}
  {Phys. Lett. B}\ }\textbf {\bibinfo {volume} {595}},\ \bibinfo {pages} {109}
  (\bibinfo {year} {2004})},\ \Eprint {http://arxiv.org/abs/hep-ex/0401004}
  {arXiv:hep-ex/0401004} \BibitemShut {NoStop}%
\bibitem [{\citenamefont {Lu}\ \emph {et~al.}(2005)\citenamefont {Lu} \emph
  {et~al.}}]{Lu:2004yn}%
  \BibitemOpen
  \bibfield  {author} {\bibinfo {author} {\bibfnamefont {M.}~\bibnamefont {Lu}}
  \emph {et~al.} (\bibinfo {collaboration} {E852}),\ }\href {\doibase
  10.1103/PhysRevLett.94.032002} {\bibfield  {journal} {\bibinfo  {journal}
  {Phys. Rev. Lett.}\ }\textbf {\bibinfo {volume} {94}},\ \bibinfo {pages}
  {032002} (\bibinfo {year} {2005})},\ \Eprint
  {http://arxiv.org/abs/hep-ex/0405044} {arXiv:hep-ex/0405044} \BibitemShut
  {NoStop}%
\bibitem [{\citenamefont {Meyer}\ and\ \citenamefont
  {Swanson}(2015)}]{Meyer:2015eta}%
  \BibitemOpen
  \bibfield  {author} {\bibinfo {author} {\bibfnamefont {C.~A.}\ \bibnamefont
  {Meyer}}\ and\ \bibinfo {author} {\bibfnamefont {E.~S.}\ \bibnamefont
  {Swanson}},\ }\href {\doibase 10.1016/j.ppnp.2015.03.001} {\bibfield
  {journal} {\bibinfo  {journal} {Prog. Part. Nucl. Phys.}\ }\textbf {\bibinfo
  {volume} {82}},\ \bibinfo {pages} {21} (\bibinfo {year} {2015})},\ \Eprint
  {http://arxiv.org/abs/1502.07276} {arXiv:1502.07276 [hep-ph]} \BibitemShut
  {NoStop}%
\bibitem [{\citenamefont {Ablikim}\ \emph
  {et~al.}(2022{\natexlab{a}})\citenamefont {Ablikim} \emph
  {et~al.}}]{BESIII:2022PRL}%
  \BibitemOpen
  \bibfield  {author} {\bibinfo {author} {\bibfnamefont {M.}~\bibnamefont
  {Ablikim}} \emph {et~al.} (\bibinfo {collaboration} {BESIII}),\ }\href@noop
  {} {\  (\bibinfo {year} {2022}{\natexlab{a}})},\ \Eprint
  {http://arxiv.org/abs/2202.00621} {arXiv:2202.00621 [hep-ex]} \BibitemShut
  {NoStop}%
\bibitem [{\citenamefont {Ablikim}\ \emph
  {et~al.}(2022{\natexlab{b}})\citenamefont {Ablikim} \emph
  {et~al.}}]{BESIII:2022PRD}%
  \BibitemOpen
  \bibfield  {author} {\bibinfo {author} {\bibfnamefont {M.}~\bibnamefont
  {Ablikim}} \emph {et~al.} (\bibinfo {collaboration} {BESIII}),\ }\href@noop
  {} {\  (\bibinfo {year} {2022}{\natexlab{b}})},\ \Eprint
  {http://arxiv.org/abs/2202.00623} {arXiv:2202.00623 [hep-ex]} \BibitemShut
  {NoStop}%
\bibitem [{\citenamefont {Chen}\ \emph {et~al.}(2022)\citenamefont {Chen},
  \citenamefont {Su},\ and\ \citenamefont {Zhu}}]{Chen:2022qpd}%
  \BibitemOpen
  \bibfield  {author} {\bibinfo {author} {\bibfnamefont {H.-X.}\ \bibnamefont
  {Chen}}, \bibinfo {author} {\bibfnamefont {N.}~\bibnamefont {Su}}, \ and\
  \bibinfo {author} {\bibfnamefont {S.-L.}\ \bibnamefont {Zhu}},\ }\href@noop
  {} {\  (\bibinfo {year} {2022})},\ \Eprint {http://arxiv.org/abs/2202.04918}
  {arXiv:2202.04918 [hep-ph]} \BibitemShut {NoStop}%
\bibitem [{\citenamefont {Qiu}\ and\ \citenamefont {Zhao}(2022)}]{Qiu:2022ktc}%
  \BibitemOpen
  \bibfield  {author} {\bibinfo {author} {\bibfnamefont {L.}~\bibnamefont
  {Qiu}}\ and\ \bibinfo {author} {\bibfnamefont {Q.}~\bibnamefont {Zhao}},\
  }\href@noop {} {\  (\bibinfo {year} {2022})},\ \Eprint
  {http://arxiv.org/abs/2202.00904} {arXiv:2202.00904 [hep-ph]} \BibitemShut
  {NoStop}%
\bibitem [{\citenamefont {Burakovsky}\ and\ \citenamefont
  {Goldman}(1997)}]{Burakovsky:1997dd}%
  \BibitemOpen
  \bibfield  {author} {\bibinfo {author} {\bibfnamefont {L.}~\bibnamefont
  {Burakovsky}}\ and\ \bibinfo {author} {\bibfnamefont {J.~T.}\ \bibnamefont
  {Goldman}},\ }\href {\doibase 10.1103/PhysRevD.56.R1368} {\bibfield
  {journal} {\bibinfo  {journal} {Phys. Rev. D}\ }\textbf {\bibinfo {volume}
  {56}},\ \bibinfo {pages} {R1368} (\bibinfo {year} {1997})},\ \Eprint
  {http://arxiv.org/abs/hep-ph/9703274} {arXiv:hep-ph/9703274} \BibitemShut
  {NoStop}%
\bibitem [{\citenamefont {Suzuki}(1993)}]{Suzuki:1993yc}%
  \BibitemOpen
  \bibfield  {author} {\bibinfo {author} {\bibfnamefont {M.}~\bibnamefont
  {Suzuki}},\ }\href {\doibase 10.1103/PhysRevD.47.1252} {\bibfield  {journal}
  {\bibinfo  {journal} {Phys. Rev. D}\ }\textbf {\bibinfo {volume} {47}},\
  \bibinfo {pages} {1252} (\bibinfo {year} {1993})}\BibitemShut {NoStop}%
\bibitem [{\citenamefont {Cheng}(2003)}]{Cheng:2003bn}%
  \BibitemOpen
  \bibfield  {author} {\bibinfo {author} {\bibfnamefont {H.-Y.}\ \bibnamefont
  {Cheng}},\ }\href {\doibase 10.1103/PhysRevD.67.094007} {\bibfield  {journal}
  {\bibinfo  {journal} {Phys. Rev. D}\ }\textbf {\bibinfo {volume} {67}},\
  \bibinfo {pages} {094007} (\bibinfo {year} {2003})},\ \Eprint
  {http://arxiv.org/abs/hep-ph/0301198} {arXiv:hep-ph/0301198} \BibitemShut
  {NoStop}%
\bibitem [{\citenamefont {Yang}(2011)}]{Yang:2010ah}%
  \BibitemOpen
  \bibfield  {author} {\bibinfo {author} {\bibfnamefont {K.-C.}\ \bibnamefont
  {Yang}},\ }\href {\doibase 10.1103/PhysRevD.84.034035} {\bibfield  {journal}
  {\bibinfo  {journal} {Phys. Rev. D}\ }\textbf {\bibinfo {volume} {84}},\
  \bibinfo {pages} {034035} (\bibinfo {year} {2011})},\ \Eprint
  {http://arxiv.org/abs/1011.6113} {arXiv:1011.6113 [hep-ph]} \BibitemShut
  {NoStop}%
\bibitem [{\citenamefont {Hatanaka}\ and\ \citenamefont
  {Yang}(2008)}]{Hatanaka:2008xj}%
  \BibitemOpen
  \bibfield  {author} {\bibinfo {author} {\bibfnamefont {H.}~\bibnamefont
  {Hatanaka}}\ and\ \bibinfo {author} {\bibfnamefont {K.-C.}\ \bibnamefont
  {Yang}},\ }\href {\doibase 10.1103/PhysRevD.77.094023} {\bibfield  {journal}
  {\bibinfo  {journal} {Phys. Rev. D}\ }\textbf {\bibinfo {volume} {77}},\
  \bibinfo {pages} {094023} (\bibinfo {year} {2008})},\ \bibinfo {note}
  {[Erratum: Phys.Rev.D 78, 059902 (2008)]},\ \Eprint
  {http://arxiv.org/abs/0804.3198} {arXiv:0804.3198 [hep-ph]} \BibitemShut
  {NoStop}%
\bibitem [{\citenamefont {Tayduganov}\ \emph {et~al.}(2012)\citenamefont
  {Tayduganov}, \citenamefont {Kou},\ and\ \citenamefont
  {Le~Yaouanc}}]{Tayduganov:2011ui}%
  \BibitemOpen
  \bibfield  {author} {\bibinfo {author} {\bibfnamefont {A.}~\bibnamefont
  {Tayduganov}}, \bibinfo {author} {\bibfnamefont {E.}~\bibnamefont {Kou}}, \
  and\ \bibinfo {author} {\bibfnamefont {A.}~\bibnamefont {Le~Yaouanc}},\
  }\href {\doibase 10.1103/PhysRevD.85.074011} {\bibfield  {journal} {\bibinfo
  {journal} {Phys. Rev. D}\ }\textbf {\bibinfo {volume} {85}},\ \bibinfo
  {pages} {074011} (\bibinfo {year} {2012})},\ \Eprint
  {http://arxiv.org/abs/1111.6307} {arXiv:1111.6307 [hep-ph]} \BibitemShut
  {NoStop}%
\bibitem [{\citenamefont {Divotgey}\ \emph {et~al.}(2013)\citenamefont
  {Divotgey}, \citenamefont {Olbrich},\ and\ \citenamefont
  {Giacosa}}]{Divotgey:2013jba}%
  \BibitemOpen
  \bibfield  {author} {\bibinfo {author} {\bibfnamefont {F.}~\bibnamefont
  {Divotgey}}, \bibinfo {author} {\bibfnamefont {L.}~\bibnamefont {Olbrich}}, \
  and\ \bibinfo {author} {\bibfnamefont {F.}~\bibnamefont {Giacosa}},\ }\href
  {\doibase 10.1140/epja/i2013-13135-3} {\bibfield  {journal} {\bibinfo
  {journal} {Eur. Phys. J. A}\ }\textbf {\bibinfo {volume} {49}},\ \bibinfo
  {pages} {135} (\bibinfo {year} {2013})},\ \Eprint
  {http://arxiv.org/abs/1306.1193} {arXiv:1306.1193 [hep-ph]} \BibitemShut
  {NoStop}%
\bibitem [{\citenamefont {Zhang}\ \emph {et~al.}(2018)\citenamefont {Zhang},
  \citenamefont {Guo},\ and\ \citenamefont {Wang}}]{Zhang:2017cbi}%
  \BibitemOpen
  \bibfield  {author} {\bibinfo {author} {\bibfnamefont {Z.-Q.}\ \bibnamefont
  {Zhang}}, \bibinfo {author} {\bibfnamefont {H.}~\bibnamefont {Guo}}, \ and\
  \bibinfo {author} {\bibfnamefont {S.-Y.}\ \bibnamefont {Wang}},\ }\href
  {\doibase 10.1140/epjc/s10052-018-5674-7} {\bibfield  {journal} {\bibinfo
  {journal} {Eur. Phys. J. C}\ }\textbf {\bibinfo {volume} {78}},\ \bibinfo
  {pages} {219} (\bibinfo {year} {2018})},\ \Eprint
  {http://arxiv.org/abs/1705.00524} {arXiv:1705.00524 [hep-ph]} \BibitemShut
  {NoStop}%
\bibitem [{\citenamefont {Roca}\ \emph {et~al.}(2005)\citenamefont {Roca},
  \citenamefont {Oset},\ and\ \citenamefont {Singh}}]{Roca:2005nm}%
  \BibitemOpen
  \bibfield  {author} {\bibinfo {author} {\bibfnamefont {L.}~\bibnamefont
  {Roca}}, \bibinfo {author} {\bibfnamefont {E.}~\bibnamefont {Oset}}, \ and\
  \bibinfo {author} {\bibfnamefont {J.}~\bibnamefont {Singh}},\ }\href
  {\doibase 10.1103/PhysRevD.72.014002} {\bibfield  {journal} {\bibinfo
  {journal} {Phys. Rev. D}\ }\textbf {\bibinfo {volume} {72}},\ \bibinfo
  {pages} {014002} (\bibinfo {year} {2005})},\ \Eprint
  {http://arxiv.org/abs/hep-ph/0503273} {arXiv:hep-ph/0503273} \BibitemShut
  {NoStop}%
\bibitem [{\citenamefont {Geng}\ \emph {et~al.}(2007)\citenamefont {Geng},
  \citenamefont {Oset}, \citenamefont {Roca},\ and\ \citenamefont
  {Oller}}]{Geng:2006yb}%
  \BibitemOpen
  \bibfield  {author} {\bibinfo {author} {\bibfnamefont {L.~S.}\ \bibnamefont
  {Geng}}, \bibinfo {author} {\bibfnamefont {E.}~\bibnamefont {Oset}}, \bibinfo
  {author} {\bibfnamefont {L.}~\bibnamefont {Roca}}, \ and\ \bibinfo {author}
  {\bibfnamefont {J.~A.}\ \bibnamefont {Oller}},\ }\href {\doibase
  10.1103/PhysRevD.75.014017} {\bibfield  {journal} {\bibinfo  {journal} {Phys.
  Rev. D}\ }\textbf {\bibinfo {volume} {75}},\ \bibinfo {pages} {014017}
  (\bibinfo {year} {2007})},\ \Eprint {http://arxiv.org/abs/hep-ph/0610217}
  {arXiv:hep-ph/0610217} \BibitemShut {NoStop}%
\bibitem [{\citenamefont {Wang}\ \emph {et~al.}(2019)\citenamefont {Wang},
  \citenamefont {Roca},\ and\ \citenamefont {Oset}}]{Wang:2019mph}%
  \BibitemOpen
  \bibfield  {author} {\bibinfo {author} {\bibfnamefont {G.~Y.}\ \bibnamefont
  {Wang}}, \bibinfo {author} {\bibfnamefont {L.}~\bibnamefont {Roca}}, \ and\
  \bibinfo {author} {\bibfnamefont {E.}~\bibnamefont {Oset}},\ }\href {\doibase
  10.1103/PhysRevD.100.074018} {\bibfield  {journal} {\bibinfo  {journal}
  {Phys. Rev. D}\ }\textbf {\bibinfo {volume} {100}},\ \bibinfo {pages}
  {074018} (\bibinfo {year} {2019})},\ \Eprint
  {http://arxiv.org/abs/1907.09188} {arXiv:1907.09188 [hep-ph]} \BibitemShut
  {NoStop}%
\bibitem [{\citenamefont {Meissner}(1988)}]{Meissner:1987ge}%
  \BibitemOpen
  \bibfield  {author} {\bibinfo {author} {\bibfnamefont {U.~G.}\ \bibnamefont
  {Meissner}},\ }\href {\doibase 10.1016/0370-1573(88)90090-7} {\bibfield
  {journal} {\bibinfo  {journal} {Phys. Rept.}\ }\textbf {\bibinfo {volume}
  {161}},\ \bibinfo {pages} {213} (\bibinfo {year} {1988})}\BibitemShut
  {NoStop}%
\bibitem [{\citenamefont {Zhang}\ \emph {et~al.}(2006)\citenamefont {Zhang},
  \citenamefont {Chiang}, \citenamefont {Shen},\ and\ \citenamefont
  {Zou}}]{Zhang:2006ix}%
  \BibitemOpen
  \bibfield  {author} {\bibinfo {author} {\bibfnamefont {Y.-J.}\ \bibnamefont
  {Zhang}}, \bibinfo {author} {\bibfnamefont {H.-C.}\ \bibnamefont {Chiang}},
  \bibinfo {author} {\bibfnamefont {P.-N.}\ \bibnamefont {Shen}}, \ and\
  \bibinfo {author} {\bibfnamefont {B.-S.}\ \bibnamefont {Zou}},\ }\href
  {\doibase 10.1103/PhysRevD.74.014013} {\bibfield  {journal} {\bibinfo
  {journal} {Phys. Rev. D}\ }\textbf {\bibinfo {volume} {74}},\ \bibinfo
  {pages} {014013} (\bibinfo {year} {2006})},\ \Eprint
  {http://arxiv.org/abs/hep-ph/0604271} {arXiv:hep-ph/0604271} \BibitemShut
  {NoStop}%
\bibitem [{\citenamefont {Dong}\ and\ \citenamefont
  {Zou}(2021)}]{Dong:2020rgs}%
  \BibitemOpen
  \bibfield  {author} {\bibinfo {author} {\bibfnamefont {X.-K.}\ \bibnamefont
  {Dong}}\ and\ \bibinfo {author} {\bibfnamefont {B.-S.}\ \bibnamefont {Zou}},\
  }\href {\doibase 10.1140/epja/s10050-021-00442-7} {\bibfield  {journal}
  {\bibinfo  {journal} {Eur. Phys. J. A}\ }\textbf {\bibinfo {volume} {57}},\
  \bibinfo {pages} {139} (\bibinfo {year} {2021})},\ \Eprint
  {http://arxiv.org/abs/2009.11619} {arXiv:2009.11619 [hep-ph]} \BibitemShut
  {NoStop}%
\bibitem [{\citenamefont {Dong}\ \emph {et~al.}(2020)\citenamefont {Dong},
  \citenamefont {Lin},\ and\ \citenamefont {Zou}}]{Dong:2019ofp}%
  \BibitemOpen
  \bibfield  {author} {\bibinfo {author} {\bibfnamefont {X.-K.}\ \bibnamefont
  {Dong}}, \bibinfo {author} {\bibfnamefont {Y.-H.}\ \bibnamefont {Lin}}, \
  and\ \bibinfo {author} {\bibfnamefont {B.-S.}\ \bibnamefont {Zou}},\ }\href
  {\doibase 10.1103/PhysRevD.101.076003} {\bibfield  {journal} {\bibinfo
  {journal} {Phys. Rev. D}\ }\textbf {\bibinfo {volume} {101}},\ \bibinfo
  {pages} {076003} (\bibinfo {year} {2020})},\ \Eprint
  {http://arxiv.org/abs/1910.14455} {arXiv:1910.14455 [hep-ph]} \BibitemShut
  {NoStop}%
\bibitem [{\citenamefont {Casalbuoni}\ \emph {et~al.}(1997)\citenamefont
  {Casalbuoni}, \citenamefont {Deandrea}, \citenamefont {Di~Bartolomeo},
  \citenamefont {Gatto}, \citenamefont {Feruglio},\ and\ \citenamefont
  {Nardulli}}]{Casalbuoni:1996pg}%
  \BibitemOpen
  \bibfield  {author} {\bibinfo {author} {\bibfnamefont {R.}~\bibnamefont
  {Casalbuoni}}, \bibinfo {author} {\bibfnamefont {A.}~\bibnamefont
  {Deandrea}}, \bibinfo {author} {\bibfnamefont {N.}~\bibnamefont
  {Di~Bartolomeo}}, \bibinfo {author} {\bibfnamefont {R.}~\bibnamefont
  {Gatto}}, \bibinfo {author} {\bibfnamefont {F.}~\bibnamefont {Feruglio}}, \
  and\ \bibinfo {author} {\bibfnamefont {G.}~\bibnamefont {Nardulli}},\ }\href
  {\doibase 10.1016/S0370-1573(96)00027-0} {\bibfield  {journal} {\bibinfo
  {journal} {Phys. Rept.}\ }\textbf {\bibinfo {volume} {281}},\ \bibinfo
  {pages} {145} (\bibinfo {year} {1997})},\ \Eprint
  {http://arxiv.org/abs/hep-ph/9605342} {arXiv:hep-ph/9605342} \BibitemShut
  {NoStop}%
\bibitem [{\citenamefont {Isola}\ \emph {et~al.}(2003)\citenamefont {Isola},
  \citenamefont {Ladisa}, \citenamefont {Nardulli},\ and\ \citenamefont
  {Santorelli}}]{Isola:2003fh}%
  \BibitemOpen
  \bibfield  {author} {\bibinfo {author} {\bibfnamefont {C.}~\bibnamefont
  {Isola}}, \bibinfo {author} {\bibfnamefont {M.}~\bibnamefont {Ladisa}},
  \bibinfo {author} {\bibfnamefont {G.}~\bibnamefont {Nardulli}}, \ and\
  \bibinfo {author} {\bibfnamefont {P.}~\bibnamefont {Santorelli}},\ }\href
  {\doibase 10.1103/PhysRevD.68.114001} {\bibfield  {journal} {\bibinfo
  {journal} {Phys. Rev. D}\ }\textbf {\bibinfo {volume} {68}},\ \bibinfo
  {pages} {114001} (\bibinfo {year} {2003})},\ \Eprint
  {http://arxiv.org/abs/hep-ph/0307367} {arXiv:hep-ph/0307367} \BibitemShut
  {NoStop}%
\bibitem [{\citenamefont {Bardeen}\ \emph {et~al.}(2003)\citenamefont
  {Bardeen}, \citenamefont {Eichten},\ and\ \citenamefont
  {Hill}}]{Bardeen:2003kt}%
  \BibitemOpen
  \bibfield  {author} {\bibinfo {author} {\bibfnamefont {W.~A.}\ \bibnamefont
  {Bardeen}}, \bibinfo {author} {\bibfnamefont {E.~J.}\ \bibnamefont
  {Eichten}}, \ and\ \bibinfo {author} {\bibfnamefont {C.~T.}\ \bibnamefont
  {Hill}},\ }\href {\doibase 10.1103/PhysRevD.68.054024} {\bibfield  {journal}
  {\bibinfo  {journal} {Phys. Rev. D}\ }\textbf {\bibinfo {volume} {68}},\
  \bibinfo {pages} {054024} (\bibinfo {year} {2003})},\ \Eprint
  {http://arxiv.org/abs/hep-ph/0305049} {arXiv:hep-ph/0305049} \BibitemShut
  {NoStop}%
\bibitem [{\citenamefont {Tornqvist}(1994)}]{Tornqvist:1993ng}%
  \BibitemOpen
  \bibfield  {author} {\bibinfo {author} {\bibfnamefont {N.~A.}\ \bibnamefont
  {Tornqvist}},\ }\href {\doibase 10.1007/BF01413192} {\bibfield  {journal}
  {\bibinfo  {journal} {Z. Phys. C}\ }\textbf {\bibinfo {volume} {61}},\
  \bibinfo {pages} {525} (\bibinfo {year} {1994})},\ \Eprint
  {http://arxiv.org/abs/hep-ph/9310247} {arXiv:hep-ph/9310247} \BibitemShut
  {NoStop}%
\bibitem [{\citenamefont {Burns}\ and\ \citenamefont
  {Swanson}(2019)}]{Burns:2019iih}%
  \BibitemOpen
  \bibfield  {author} {\bibinfo {author} {\bibfnamefont {T.~J.}\ \bibnamefont
  {Burns}}\ and\ \bibinfo {author} {\bibfnamefont {E.~S.}\ \bibnamefont
  {Swanson}},\ }\href {\doibase 10.1103/PhysRevD.100.114033} {\bibfield
  {journal} {\bibinfo  {journal} {Phys. Rev. D}\ }\textbf {\bibinfo {volume}
  {100}},\ \bibinfo {pages} {114033} (\bibinfo {year} {2019})},\ \Eprint
  {http://arxiv.org/abs/1908.03528} {arXiv:1908.03528 [hep-ph]} \BibitemShut
  {NoStop}%
\bibitem [{\citenamefont {Yalikun}\ \emph {et~al.}(2021)\citenamefont
  {Yalikun}, \citenamefont {Lin}, \citenamefont {Guo}, \citenamefont {Kamiya},\
  and\ \citenamefont {Zou}}]{Yalikun:2021bfm}%
  \BibitemOpen
  \bibfield  {author} {\bibinfo {author} {\bibfnamefont {N.}~\bibnamefont
  {Yalikun}}, \bibinfo {author} {\bibfnamefont {Y.-H.}\ \bibnamefont {Lin}},
  \bibinfo {author} {\bibfnamefont {F.-K.}\ \bibnamefont {Guo}}, \bibinfo
  {author} {\bibfnamefont {Y.}~\bibnamefont {Kamiya}}, \ and\ \bibinfo {author}
  {\bibfnamefont {B.-S.}\ \bibnamefont {Zou}},\ }\href {\doibase
  10.1103/PhysRevD.104.094039} {\bibfield  {journal} {\bibinfo  {journal}
  {Phys. Rev. D}\ }\textbf {\bibinfo {volume} {104}},\ \bibinfo {pages}
  {094039} (\bibinfo {year} {2021})},\ \Eprint
  {http://arxiv.org/abs/2109.03504} {arXiv:2109.03504 [hep-ph]} \BibitemShut
  {NoStop}%
\bibitem [{\citenamefont {Weinberg}(1965)}]{Weinberg:1965zz}%
  \BibitemOpen
  \bibfield  {author} {\bibinfo {author} {\bibfnamefont {S.}~\bibnamefont
  {Weinberg}},\ }\href {\doibase 10.1103/PhysRev.137.B672} {\bibfield
  {journal} {\bibinfo  {journal} {Phys. Rev.}\ }\textbf {\bibinfo {volume}
  {137}},\ \bibinfo {pages} {B672} (\bibinfo {year} {1965})}\BibitemShut
  {NoStop}%
\bibitem [{\citenamefont {Baru}\ \emph {et~al.}(2004)\citenamefont {Baru},
  \citenamefont {Haidenbauer}, \citenamefont {Hanhart}, \citenamefont
  {Kalashnikova},\ and\ \citenamefont {Kudryavtsev}}]{Baru:2003qq}%
  \BibitemOpen
  \bibfield  {author} {\bibinfo {author} {\bibfnamefont {V.}~\bibnamefont
  {Baru}}, \bibinfo {author} {\bibfnamefont {J.}~\bibnamefont {Haidenbauer}},
  \bibinfo {author} {\bibfnamefont {C.}~\bibnamefont {Hanhart}}, \bibinfo
  {author} {\bibfnamefont {Y.}~\bibnamefont {Kalashnikova}}, \ and\ \bibinfo
  {author} {\bibfnamefont {A.~E.}\ \bibnamefont {Kudryavtsev}},\ }\href
  {\doibase 10.1016/j.physletb.2004.01.088} {\bibfield  {journal} {\bibinfo
  {journal} {Phys. Lett. B}\ }\textbf {\bibinfo {volume} {586}},\ \bibinfo
  {pages} {53} (\bibinfo {year} {2004})},\ \Eprint
  {http://arxiv.org/abs/hep-ph/0308129} {arXiv:hep-ph/0308129} \BibitemShut
  {NoStop}%
\bibitem [{\citenamefont {Baru}\ \emph {et~al.}(2011)\citenamefont {Baru},
  \citenamefont {Filin}, \citenamefont {Hanhart}, \citenamefont {Kalashnikova},
  \citenamefont {Kudryavtsev},\ and\ \citenamefont {Nefediev}}]{Baru:2011rs}%
  \BibitemOpen
  \bibfield  {author} {\bibinfo {author} {\bibfnamefont {V.}~\bibnamefont
  {Baru}}, \bibinfo {author} {\bibfnamefont {A.~A.}\ \bibnamefont {Filin}},
  \bibinfo {author} {\bibfnamefont {C.}~\bibnamefont {Hanhart}}, \bibinfo
  {author} {\bibfnamefont {Y.~S.}\ \bibnamefont {Kalashnikova}}, \bibinfo
  {author} {\bibfnamefont {A.~E.}\ \bibnamefont {Kudryavtsev}}, \ and\ \bibinfo
  {author} {\bibfnamefont {A.~V.}\ \bibnamefont {Nefediev}},\ }\href {\doibase
  10.1103/PhysRevD.84.074029} {\bibfield  {journal} {\bibinfo  {journal} {Phys.
  Rev. D}\ }\textbf {\bibinfo {volume} {84}},\ \bibinfo {pages} {074029}
  (\bibinfo {year} {2011})},\ \Eprint {http://arxiv.org/abs/1108.5644}
  {arXiv:1108.5644 [hep-ph]} \BibitemShut {NoStop}%
\bibitem [{\citenamefont {Du}\ \emph {et~al.}(2022)\citenamefont {Du},
  \citenamefont {Baru}, \citenamefont {Dong}, \citenamefont {Filin},
  \citenamefont {Guo}, \citenamefont {Hanhart}, \citenamefont {Nefediev},
  \citenamefont {Nieves},\ and\ \citenamefont {Wang}}]{Du:2021zzh}%
  \BibitemOpen
  \bibfield  {author} {\bibinfo {author} {\bibfnamefont {M.-L.}\ \bibnamefont
  {Du}}, \bibinfo {author} {\bibfnamefont {V.}~\bibnamefont {Baru}}, \bibinfo
  {author} {\bibfnamefont {X.-K.}\ \bibnamefont {Dong}}, \bibinfo {author}
  {\bibfnamefont {A.}~\bibnamefont {Filin}}, \bibinfo {author} {\bibfnamefont
  {F.-K.}\ \bibnamefont {Guo}}, \bibinfo {author} {\bibfnamefont
  {C.}~\bibnamefont {Hanhart}}, \bibinfo {author} {\bibfnamefont
  {A.}~\bibnamefont {Nefediev}}, \bibinfo {author} {\bibfnamefont
  {J.}~\bibnamefont {Nieves}}, \ and\ \bibinfo {author} {\bibfnamefont
  {Q.}~\bibnamefont {Wang}},\ }\href {\doibase 10.1103/PhysRevD.105.014024}
  {\bibfield  {journal} {\bibinfo  {journal} {Phys. Rev. D}\ }\textbf {\bibinfo
  {volume} {105}},\ \bibinfo {pages} {014024} (\bibinfo {year} {2022})},\
  \Eprint {http://arxiv.org/abs/2110.13765} {arXiv:2110.13765 [hep-ph]}
  \BibitemShut {NoStop}%
\bibitem [{\citenamefont {Guo}\ and\ \citenamefont
  {Meissner}(2011)}]{Guo:2011dd}%
  \BibitemOpen
  \bibfield  {author} {\bibinfo {author} {\bibfnamefont {F.-K.}\ \bibnamefont
  {Guo}}\ and\ \bibinfo {author} {\bibfnamefont {U.-G.}\ \bibnamefont
  {Meissner}},\ }\href {\doibase 10.1103/PhysRevD.84.014013} {\bibfield
  {journal} {\bibinfo  {journal} {Phys. Rev. D}\ }\textbf {\bibinfo {volume}
  {84}},\ \bibinfo {pages} {014013} (\bibinfo {year} {2011})},\ \Eprint
  {http://arxiv.org/abs/1102.3536} {arXiv:1102.3536 [hep-ph]} \BibitemShut
  {NoStop}%
\bibitem [{\citenamefont {Molina}\ \emph {et~al.}(2008)\citenamefont {Molina},
  \citenamefont {Nicmorus},\ and\ \citenamefont {Oset}}]{Molina:2008jw}%
  \BibitemOpen
  \bibfield  {author} {\bibinfo {author} {\bibfnamefont {R.}~\bibnamefont
  {Molina}}, \bibinfo {author} {\bibfnamefont {D.}~\bibnamefont {Nicmorus}}, \
  and\ \bibinfo {author} {\bibfnamefont {E.}~\bibnamefont {Oset}},\ }\href
  {\doibase 10.1103/PhysRevD.78.114018} {\bibfield  {journal} {\bibinfo
  {journal} {Phys. Rev. D}\ }\textbf {\bibinfo {volume} {78}},\ \bibinfo
  {pages} {114018} (\bibinfo {year} {2008})},\ \Eprint
  {http://arxiv.org/abs/0809.2233} {arXiv:0809.2233 [hep-ph]} \BibitemShut
  {NoStop}%
\bibitem [{\citenamefont {Woss}\ \emph {et~al.}(2021)\citenamefont {Woss},
  \citenamefont {Dudek}, \citenamefont {Edwards}, \citenamefont {Thomas},\ and\
  \citenamefont {Wilson}}]{Woss:2020ayi}%
  \BibitemOpen
  \bibfield  {author} {\bibinfo {author} {\bibfnamefont {A.~J.}\ \bibnamefont
  {Woss}}, \bibinfo {author} {\bibfnamefont {J.~J.}\ \bibnamefont {Dudek}},
  \bibinfo {author} {\bibfnamefont {R.~G.}\ \bibnamefont {Edwards}}, \bibinfo
  {author} {\bibfnamefont {C.~E.}\ \bibnamefont {Thomas}}, \ and\ \bibinfo
  {author} {\bibfnamefont {D.~J.}\ \bibnamefont {Wilson}} (\bibinfo
  {collaboration} {Hadron Spectrum}),\ }\href {\doibase
  10.1103/PhysRevD.103.054502} {\bibfield  {journal} {\bibinfo  {journal}
  {Phys. Rev. D}\ }\textbf {\bibinfo {volume} {103}},\ \bibinfo {pages}
  {054502} (\bibinfo {year} {2021})},\ \Eprint
  {http://arxiv.org/abs/2009.10034} {arXiv:2009.10034 [hep-lat]} \BibitemShut
  {NoStop}%
\end{thebibliography}%

\end{document}